\documentclass[11pt,a4paper]{article} 
 
        \usepackage[utf8]{inputenc}
        \usepackage{fontenc}
	\usepackage[italian,english]{babel}
	\usepackage{amsmath}
	\usepackage{geometry}
	\usepackage{amsthm}
	\usepackage{mathrsfs}
	\usepackage{bbold}
	\usepackage{mathtools}
	\usepackage{amsfonts}
	\usepackage{graphicx}
	\usepackage{physics}
	 \usepackage{comment}
	 \usepackage{bbm}
	  \usepackage{layout}
	\usepackage{xcolor}
     \usepackage[big]{layaureo}
	 \usepackage{dsfont}	
	 \usepackage{slashed}
 	 \usepackage{amssymb}
	 \usepackage{tensor}
	\usepackage{fancyhdr} 
 	\usepackage{tikz-cd}
 	\usepackage{braket}
 	\usepackage{tikz}
 	\usepackage[compat=1.1.0]{tikz-feynman}
 	\usepackage{subfigure}
	\usepackage{fancyhdr} 
	\usepackage{booktabs}
	\usepackage{braket}
	
\usepackage{tikz}

\usetikzlibrary{
	decorations.markings,
	decorations.pathmorphing,
}

\def\graphwidth{3.0}
\def\graphheight{1.2}

\def\contourgap{0.12}
\def\axescolor{black}
\def\contourcolor{red}
\def\endfirstcut{0.75}
\def\beginsecondcut{2.25}
\tikzset{cutstyle/.style={decorate, decoration={zigzag, segment length=6, amplitude=2}, draw=black}}
\tikzset{arrow data/.style 2 args={decoration={markings, mark=at position #1 with \arrow{#2}}, postaction=decorate}}
\tikzset{partial ellipse/.style args={#1:#2:#3}{insert path={+ (#1:#3) arc (#1:#2:#3)}}}

\newcommand{\drawaxes}[2]{
	\draw[-latex, \axescolor] (-\graphwidth/2, 0) -- (+\graphwidth, 0) node[right] {#1};
	\draw[-latex, \axescolor] (0, -\graphheight) -- (0, +\graphheight) node[below left] {#2};
}
	        
\usepackage{jheppub} 

\usepackage{natbib}
\title{The resurgence of the plateau in supersymmetric  ${\cal N}=1$ Jackiw-Teitelboim gravity}

	\newcommand{\beq}{\begin{equation}}
	\newcommand{\bea}{\begin{eqnarray}}
	\newcommand{\eea}{\end{eqnarray}}
	\newcommand{\eeq}{\end{equation}}

	\newcommand{\de}{\partial}

	\makeatletter
	\newcommand{\aextp}{\@ifnextchar^\@aextp{\@aextp^{\,}}}
	\def\@aextp^#1{\mathop{\bigwedge\nolimits^{\!#1}}}
	\makeatother
	
	\makeatletter
	\newcommand{\extp}{\@ifnextchar_\@extp{\@extp_{\,}}}
	\def\@extp_#1{\mathop{\aextp\nolimits_{\!#1}}}
	\makeatother

	\theoremstyle{definition}

\author[a]{Luca Griguolo,}
\author[b]{Jacopo Papalini,}
\author[c]{Lorenzo Russo,}
\author[c]{and Domenico Seminara}

\affiliation[a]{Dipartimento SMFI, Universit\`a di Parma and INFN Gruppo Collegato di Parma,
	Viale G.P. Usberti 7/A, 43100 Parma, Italy}
\affiliation[b]{Galileo Galilei Institute for Theoretical Physics, INFN,
	Largo Enrico Fermi, 2, 50125 Firenze, Italy}
\affiliation[c]{Dipartimento di Fisica, Universit\`a di Firenze and INFN Sezione di Firenze, via G. Sansone 1, 50019 Sesto Fiorentino, Italy} 

\emailAdd{luca.griguolo@unipr.it}
\emailAdd{jacopo.papalini@unipr.it}
\emailAdd{lorenzo.russo@unifi.it}
\emailAdd{seminara@fi.infn.it}

\abstract{Significant progresses have been made recently in understanding the spectral form factor of Jackiw-Teitelboim gravity, particularly at late times where non-perturbative effects are expected to play a dominant role. By focusing on a peculiar regime of large time and fixed temperature, called $\tau$-scaling limit, it was found that it is possible to analytically investigate the late-time plateau directly through the gravitational genus expansion. We extend this analysis to the supersymmetric  $\mathcal{N}=1$ generalization of the bosonic theory, revealing an interesting structure. First, we notice that in the $\tau$-scaling limit the perturbative sum over genera truncates after a single term, which solely accounts for the ramp behaviour. Instead a non-perturbative completion, responsible for the plateau, is encoded into an exact formula coming from the properties of the chiral gaussian ensemble, governing the spectral properties of the supersymmetric theory. We are able to recover the non-perturbative contributions by slightly deforming the genus of the involved surfaces and using resurgence theory. We derive a closed-form analytical expression for the late-time plateau and a trans-series expansion that captures the ramp-plateau transition.}

\begin{document}
\maketitle
\flushbottom

\section{Introduction}
An important challenge in contemporary theoretical physics is to explain the discrete spectrum of microstates of black holes, as predicted by a complete theory of quantum gravity. In recent years, significant progress has been made in this direction, mainly exploiting non-perturbative quantum properties of low-dimensional gravities. In this context wormholes geometries have played a prominent role:  they enter in the derivation of the Page curve \cite{Almheiri_2020, Penington:2019kki} and elucidate various aspects related to late-time correlators  \cite{Blommaert_2019, Saad:2019pqd,  Blommaert:2020seb}. Wormholes also appear in direct analysis of the discreteness of energy levels of quantum black holes. A notable example is the spectral form factor (SFF), a toy model for the thermal two-point function which is particularly valuable in measuring level statistics in quantum chaotic systems \cite{PhysRevLett.56.2449}.

In a dynamical theory of gravity, Maldacena \cite{Maldacena:2001kr} pointed out that correlation functions measured in black hole backgrounds should not decay at late times, as black holes themselves represent finite entropy systems. Consequently, the behavior of the SFF serves as an excellent diagnostic for discerning the discreteness of energy levels \cite{Cotler:2016fpe}. A concrete application of these ideas, in the context of quantum gravity and holography, is the inspection of the SFF in the Sachdev-Ye-Kitaev (SYK) model \cite{Sachdev:1992fk,Kitaev:2017awl}: this observable displays a peculiar ramp and plateau structure as a function of time \cite{Cotler:2016fpe,You_2017,Garcia-Garcia:2016mno} and supports the conjecture that statistical properties of the energy levels of quantum chaotic systems are universally described by random matrix models \cite{Bohigas:1983er}. As a matter of fact, Jackiw-Teitelboim (JT) gravity \cite{Jackiw:1984je, Teitelboim:1983ux}, which is holographically dual to the low-energy sector of the SYK model was later found to be described by a double-scaled random matrix model \cite{Saad:2019lba}. In JT gravity the form of the SFF in not completely understood from the gravitational perspective, as the ramp arises from the perturbative contribution of a wormhole \cite{Saad:2018bqo} connecting the two boundaries of the spacetime, but the plateau necessitates of additional non-perturbative effects \cite{Cotler:2016fpe, Saad:2019lba}. Instead, the complementary picture from the matrix model side is more encouraging and in \cite{Johnson:2020exp} it was demonstrated how the correct structure of the SFF, incorporating essential non-perturbative contributions, can be numerically obtained by interpreting JT gravity as a combination of minimal string models.

Nonetheless, it is of utmost importance to gain some insights into the physics that governs the transition from the ramp to the plateau directly from a bulk geometrical perspective. Remarkably, it was recently realized \cite{Saad:2022kfe, Blommaert:2022lbh, Okuyama:2023pio} that the perturbative topological expansion of JT gravity contains far more information than initially believed. By taking a suitable rescaling for the late time, known as the $\tau$-scaling limit, it becomes possible to sum the entire asymptotic perturbative series and to obtain an expression that not only captures the ramp but also accounts for the plateau. 

Encouraged by these results, we perform here the analysis of the SFF in the $\mathcal{N}=1$ Super JT gravity (SJT), a supersymmetric generalization of the bosonic theory \cite{Stanford:2019vob}, in the $\tau$-scaling limit. Previous studies of the SFF in this theory were conducted in \cite{Johnson:2020exp} exploiting RMT tools, an approach that required significant numerical efforts. We argue that it becomes instead possible to investigate the plateau behavior of the theory analytically, thereby highlighting the form of the necessary non-perturbative contributions that complement the gravitational topological expansion.

\paragraph{Summary of results} The paper is organized as follows.
In Section \ref{tau_scaling}, we review the main results concerning the $\tau$-scaling limit of the SFF in JT gravity and its low-energy regime, the Airy model. By exploiting the universal behavior of the density-density correlator in matrix models of the unitary class, we obtain an integral representation for the $\tau$-scaled SFF and demonstrate that the bulk topological expansion becomes convergent. Interestingly, after the summation, the result features a ramp and a late-time plateau, allowing for the investigation of seemingly non-perturbative phenomena directly from the bulk.

In Section \ref{bulk} we explore the correspondent genus expansion for the $\tau$-scaled SFF in SJT gravity. With a simple proof based on the mathematical properties of the Weyl-Petersson super-volumes, we show that only the genus zero term contributes to our observable. In detail, only a linear piece in $\tau$ survives. This term arises from the cylinder surface built by gluing two asymptotic trumpets with all the corrections coming from the surfaces of higher genus vanishing. Using this result, we argue that in SJT gravity only the ramp can be accessed perturbatively and it becomes imperative to incorporate certain non-perturbative effects to witness the emergence of the plateau.

Guided by this observation, in Section \ref{Bessel} we begin the investigation of the non-perturbative structure of the SFF of SJT gravity drawing upon the correspondence of the theory to an Altland-Zirnbauer matrix model, the so called \textit{chiral Gaussian unitary ensemble}. The statistics governing the correlation between close eigenvalues here has been recently studied in \cite{Akemann:2021uhf}: as long as the eigenvalues are in the bulk of the spectral support, the repulsion between them is governed by the universal sine kernel, appearing also in the bosonic case. On the other hand, as one moves towards the hard edge of the spectrum, the eigenvalue statistics is controlled by the Bessel kernel, that controls the low-energy limit of SJT. Moreover, in \cite{Akemann:2021uhf} it has been shown that, as the level spacing goes to zero, the Bessel kernel becomes well approximated by the sine kernel. Relying on this finding, we argue that the integral representation, obtained in the previous section, can be employed to compute the $\tau$-scaled SFF for both the Bessel model and SJT gravity.  By explicitly evaluating the observable in the case of the Bessel model, we find that the perturbative series in $\tau$ terminates after the linear term, as expected from the genus expansion. However, a non-perturbative completion emerges, precisely determining the late-time plateau.

The remaining part of the paper is dedicated to deduce the analytic form of these non-perturbative contributions in SJT gravity from a more geometrical perspective. More precisely, our objective is to obtain a trans-series representation for the $\tau$-scaled SFF relying on the techniques of resurgence theory \cite{ecalle1985fonctions}. This approach provides a general mathematical framework to extract the non-perturbative pieces necessary to have a well-defined physical observable solely in terms of perturbative data. Resurgence techniques have been already applied to JT gravity: we stress that without the $\tau$-scaling limit the genus expansion is in fact asymptotic and resurgence is an ideal approach to recover its non-perturbative completion. In a beautiful series of papers \cite{Gregori:2021tvs,Eynard:2023qdr,Schiappa:2023ned} the resurgent nature of JT gravity has been carefully studied, starting both from its minimal string embedding and its Weil-Peterson volumes interpretation. As a result instanton-actions and Stokes data for transseries have been computed and tested, while the calculations have been systematized into the framework of a non-perturbative extension of topological recursion relation \cite{Eynard:2007kz}. Remarkably, in the present case, only the perturbative contributions have a clear geometrical interpretation and it would be insightful to derive some links between the genus expansion and the non-perturbative matrix model results. As already noticed, in SJT gravity only one perturbative coefficient is non-vanishing in the $\tau$-scaling limit and it would seem impossible to successfully complete the resurgence program, that is usually based on the properties of asymptotic series. Our central observation is that the perturbative contribution to the SSF is encoded in a particular coefficient $P_g^{(\rho)}(\beta)$, weighting the contribution of a genus $g$ topology and depending on the original spectral density $\rho$. In Section \ref{contour} we find that by formally deforming the parameter $g$, the perturbative series in $\tau$ defined through $P_g^{(\rho)}(\beta)$ immediately becomes asymptotic and admits a resurgent analysis. This particular realization of resurgence theory is known in the literature as Cheshire cat resurgence \cite{article1, Dorigoni:2017smz, Dorigoni:2019kux}, and it has been already exploited in previous papers to complete models with "trivial" perturbative series \cite{Fujimori:2022qij}. Thanks to this method we are able to recover the exact form of the non-perturbative correction in the low-energy regime of SJT gravity, i.e. the Bessel model, fixing the Stokes coefficient that could not be determined by resurgence alone.

Leveraging this result, in Section \ref{simplified}, we investigate the SFF in a mathematical example that interpolates between the Bessel and the Airy models. This artificial construction, emerging from the low-energy expansion of the SJT spectral density, not only serves as a toy model for the most general case but will also provide a generating function for the computation of the SFF in SJT gravity. Using our method we are able to determine a trans-series expression for the $\tau$-scaled SFF and determine an analytic form for its late-time plateau. We also perform a numerical analysis highlighting a remarkable agreement between the trans-series and the exact result.

In Section \ref{SJT} we finally generalize the analysis to the theory of SJT gravity. In this case, the technical part of the computation is much more intricate, but we are still able to obtain a trans-series expression for the $\tau$-scaled SFF. Our conclusion are contained in Section \ref{frodeno}, while we defer in Appendix \ref{degregori} some technical aspects of our computations.

\section{Review of $\tau$-scaling limit of $\mathrm{SFF}$}\label{tau_scaling}
In this section, we will briefly summarize the main results of the SFF in the $\tau$-scaling limit.
So far, at least to our knowledge, this analysis was carried out for the JT gravity case and its low energy limit, the Airy model \cite{Saad:2022kfe,Okuyama:2023pio}. Interestingly, it was discovered that the spectral form factor for these two theories exhibits a simple form with a convergent genus expansion \cite{Saad:2022kfe}. 

The spectral form factor is defined as the expectation value of the product of two  analytically continued partition functions, i.e. 
\begin{equation}
\label{SSF}
\mathrm{SSF}(\beta,t)\equiv \langle Z(\beta + it) Z(\beta - it) \rangle,
 \end{equation}
where $Z(\beta) = \Tr e^{-\beta H}$. We can express the r.h.s of \eqref{SSF}  as a double Laplace transform over the energies $E_1$ and $E_2$
of the pair-correlation density\footnote{We recall that $\rho(E_1,E_2)=\expval{\tilde\rho(E_1)\tilde\rho(E_2)}$, where $\tilde\rho(E)=\sum_k\delta(E-E_k) $ is the spectral density.}$\rho(E_1,E_2)$:
\begin{equation}\label{ZZ}
\mathrm{SSF}(\beta,t) = \int_{0}^{\infty} \mathrm{d}E_1 \int_{0}^{\infty} \mathrm{d} E_2 e^{-\beta(E_1 +E_2)} e^{-i t(E_1 - E_2)} \rho(E_1, E_2).
\end{equation}
Below, we shall focus on the connected part of the $\mathrm{SSF}(\beta,t)$, which is defined as 
\[
\mathrm{SSF}_{\rm c}(\beta,t) \equiv
\langle Z(\beta + it) Z(\beta - it)  \rangle _{c} = \langle Z(\beta + it) Z(\beta - it) \rangle - \expval{Z(\beta + it)}\expval{ Z(\beta - it)}.
\]
In particular, we are interested in its behavior at late times. Using the stationary phase method for large values of $t$, the integral \eqref{ZZ}  is dominated by contributions from regions where $E_1$ and $E_2$ are very close to each other. In this regime  we can express $\mathrm{SSF}_{\rm c}(\beta,t)$ as
\begin{equation}
\mathrm{SSF}_{\rm c}(\beta,t) \underset{t\rightarrow +\infty}{\approx} \int_{0}^{\infty} \mathrm{d}E_1 \int_{0}^{\infty} \mathrm{d} E_2 e^{-\beta(E_1 + E_2)} e^{it(E_1 - E_2)} \rho_{\rm c}^{\mathrm{eff}}(E_1, E_2)
\end{equation}
where $\rho_{\rm c}^{\mathrm{eff}}(E_1, E_2)$ is an effective pair-correlation density experienced by neighboring energy levels. For the cases under scrutiny, it is given by the well-known sine kernel:
\begin{equation}\label{sine}
\rho_{\rm c}^{\mathrm{eff}}(E_1, E_2)= \delta \left(E_{-} \right) \rho (E) - \frac{ \sin^2 \left( \pi \rho(E) E_- \right)}{\pi^2 E_-^2},
\end{equation}
where $E_-\equiv E_1-E_2$ and $E\equiv\frac{E_1+E_2}{2}$.
The result \eqref{sine} actually holds for any random matrix potential $V(H)$ in the unitary universality class\footnote{This is a direct manifestation of random matrix universality, governing universal level statistics at small energy separations.}.

Performing the integration over $E_-$ in \eqref{sine} in the large $t-$limit  leads to the following integral representation of  $\mathrm{SSF}_{\rm c}(\beta,t)$ in terms of the usual  eigenvalue density $\rho(E)$:  \begin{equation}\label{SFF}
\mathrm{SSF}_{\rm c}(\beta,t) \underset{t\rightarrow +\infty}{\approx} \int_{0}^{\infty} \mathrm{d} E e^{-2 \beta E} \min\left(\rho (E), \frac{t}{2\pi}\right)
\end{equation}
In \cite{Saad:2022kfe} it was argued that the {\it uncontrolled} approximation \eqref{SFF}  can be made rigorous  and ultimately exact in the so-called 
$\tau$-scaled limit. It is a double-scaling limit where $t$ and $S_0$ go to $\infty$  while $\tau=e^{-S_{0}} t$ is held fixed. Then the spectral form factor in this limit, $\mathrm{SFF}_{\infty}(\beta,\tau)$ is defined as
\begin{equation}
\mathrm{SFF}_{\infty}(\beta,\tau)\equiv\lim_{S_0 \to +\infty} e^{-S_0} \langle Z(\beta + i\tau e^{S_0}) Z(\beta - i\tau e^{S_0}) \rangle_{\rm c}
\end{equation}
and it is computed by the r.h.s. of \eqref{SFF}. This fact was first checked for the Airy model  \cite{Saad:2022kfe}, which provides the low energy approximation of JT gravity on orientable Riemann surfaces. In fact, if one exploits the disk-level spectral density $\rho_{\rm Airy} (E) = \frac{\sqrt{E}}{2\pi}$, the integral \eqref{SFF} yields:
\begin{equation}\label{perturbative_plateau}
\begin{split}
\mathrm{SFF}_{\infty}^{\mathrm{Airy}}(\beta,\tau)&= \int_{0}^{\infty} \mathrm{d} E e^{-2 \beta E} \min\left(\frac{\sqrt{E}}{2\pi}, \frac{\tau}{2\pi}\right)\\ 
&= \frac{1}{2\pi}\frac{\pi^{\frac{1}{2}}}{2^{\frac{5}{2}}\beta^{\frac{3}{2}}} \erf\left(\sqrt{2\beta}\tau \right) = \frac{\tau}{4\beta\pi} - \frac{\tau^3}{6\pi} + \frac{\beta}{10\pi}\tau^5+...
\end{split}\end{equation}
Strikingly, this result perfectly reproduces the $\tau$-scaled limit evaluated from the exact expression of the spectral form factor within the Airy model  \cite{Okuyama:2021cub}.  Furthermore, an insightful interpretation of the terms appearing in the expansion \eqref{perturbative_plateau} in terms of a semiclassical theory of ``encounters'' of periodic orbits was discussed in \cite{Saad:2022kfe}. 

When applied to JT gravity, whose spectral density is  $\rho_{\rm JT} (E) = \frac{1}{4\pi^2} \sinh \left(2 \pi \sqrt{E}\right)$, the integral representation \eqref{SFF}  of the spectral form factor in the $\tau$-scaled limit yields
\begin{equation}\label{JT}
\begin{split}
\mathrm{SFF}_{\infty}^{\mathrm{JT}}&(\beta,\tau)=\frac{1}{2 \pi} \int_{0}^{\infty} \mathrm{d} E  e^{-2 \beta E} \min\left(\frac{1}{2\pi} \sinh \left(2 \pi \sqrt{E}\right), \tau\right) \\
&=\frac{e^{\frac{\pi^2}{2 \beta}}}{16 \sqrt{2 \pi}\beta^{\frac32}} \left[\erf\left(\frac{\frac{\beta}{\pi}\mathrm{arcsinh}(2\pi\tau)+\pi}{\sqrt{2\beta}}\right)+\erf\left(\frac{\frac{\beta}{\pi}\mathrm{arcsinh}(2\pi\tau)-\pi}{\sqrt{2\beta}}\right) \right].
\end{split}
\end{equation}
While the usual genus expansion of the complete SFF in JT gravity is an asymptotic series, expanding \eqref{JT} in powers of $\tau$ produces a convergent series\footnote{One can immediately realize the series in $\tau$ corresponds to the genus expansion, by undoing the $\tau$ scaling.}.  This result was significant, as it paved the way for a perturbative treatment of the late-time plateau. Unlike the Airy model, an exact closed expression for the SFF in
 JT gravity remains elusive. However one can check the validity of the result \eqref{JT} using the topological recursion techniques. In \cite{Saad:2022kfe}   it was verified that the expansion of \eqref{JT} accurately reproduces the $\tau$-scaled genus expansion up to order $\tau^{13}$.

Additionally, in cases where the spectral density is a monotonically increasing function of energy, as occurs in JT gravity and the Airy model, the coefficient of $\tau^{2g+1}$ for $g\geq 1$ in the Taylor-expansion of the $\tau-$scaled spectral form factor can be computed using the subsequent formula:
\begin{equation}\label{Pn}
P_g^{(\rho)}(\beta)=- \frac{1}{g (2g +1)(2\pi)^{2g+1}}\oint \frac{\mathrm{d}E}{2 \pi i} \frac{e^{-2 \beta E}}{\rho_0(E)^{2g}}.
\end{equation}
Here the integral is evaluated on a contour around the origin. In  \cite{Saad:2022kfe} it was demonstrated that this prescription, upon summation, reproduces the integral representation \eqref{SFF}, i.e. 
\begin{equation}\label{summation}
\begin{split}
\mathrm{SFF}(\beta, \tau)&= \frac{\tau}{4\pi \beta}+ \sum_{g=1}^{\infty}  P_g^{(\rho)}(\beta) \tau^{2g +1} =\int_{0}^{\infty} \mathrm{d} E e^{-2 \beta E} \min\left(\rho_{0}(E), \frac{\tau}{2\pi}\right).
\end{split}
\end{equation}
The aim of the upcoming sections is to broaden the application of this formalism within the framework of SJT gravity. A pivotal technical step in achieving this goal will entail extending the definition of \eqref{Pn} to accommodate a non-monotonically increasing spectral density.

\section{The $\tau$-scaling limit for SJT gravity}
\subsection{A perturbative bulk expansion}\label{bulk}

This section is devoted to understanding the general form of all terms in the genus expansion of the Spectral Form Factor in SJT gravity\footnote{In this paper we only consider the $\mathcal{N}=1$ supersymmetric generalization of the bosonic theory where   time reversal is  not gauged and each spin structure of the manifolds appearing in the path integral is weighted by $(-1)^{\zeta}$, with $\zeta$ the Atiyah-Singer index. } from the bulk point of view. This derivation closely mirrors the one conducted in \cite{Blommaert:2022lbh} for JT gravity. While in the cases of JT gravity, the cancellation of certain terms exhibiting unfavorable growth with respect to $\tau$ in the genus $g$ allows for perturbative access to the plateau during the $\tau$-scaling limit, our findings in the SJT scenario reveal a different outcome. Specifically, we observe that all higher genus terms vanish in the $\tau$-scaling limit.

By employing the standard topological decomposition of the gravitational path integral, the key ingredients for our analysis consist of the volumes of moduli space of super Riemann surfaces of genus $g$ and $n$ boundaries having lengths $b_1,\cdots,b_n$. These volume are denotes as $\tilde{V}_{g,n}(b_1, ..., b_n)$. Furthermore we need the partition function $Z_{\mathrm{SJT}}^{\mathrm{tr}}(\beta, b)$  on the \textit{trumpet} spacetime.  In fact, these two objects serve as the building blocks to construct the partition function on any topology through a simple gluing procedure.

The partition function of SJT on the basic topologies, namely the disk and the trumpet, was explicitly computed in [22] using the super-Schwarzian model of the boundary theory. The expressions for these partition functions are as follows:
\begin{equation}\label{basic}
Z_{\mathrm{SJT}}^{\mathrm{disk}}(\beta, b) = \sqrt{\frac{2}{\pi \beta}}e^{\frac{\pi^2}{\beta}}  \qquad Z_{\mathrm{SJT}}^{\mathrm{tr}}(\beta, b) = \frac{1}{\sqrt{2 \pi \beta}} e^{-\frac{b^2}{4\beta}} 
\end{equation}
To investigate the SFF, we have to compute the connected two-boundary correlator of the theory.  This correlator exhibits the following genus expansion: 
\begin{equation}\label{series}
Z^{\mathrm{SJT}}(\beta_1, \beta_2)= \sum_{g=0}^{+\infty} Z_{g}^{\mathrm{SJT}}(\beta_1, \beta_2)+[\mathrm{nonperturbative \ terms}].
\end{equation}
The investigation of the possible nonperturbative completion appearing in the expansion \eqref{series} will constitute the topic of the next Section. In \eqref{series}
we denoted with $Z_{g}^{\mathrm{SJT}}(\beta_1, \beta_2)$ the genus-g gravitational wormhole amplitude that can be evaluated by gluing together two trumpets through the insertion of a  Weil-Petersson volume $\tilde{V}_{g, 2}(b_1, b_2)$. Achieving this involves  integrating over all possible lengths, $b_1$ and $b_2$, of the geodesic boundaries, i.e.
\begin{equation}\label{wormhole}
Z_g^{\mathrm{SJT}}(\beta_1, \beta_2) = e^{-2g S_0} \int_{0}^{\infty} b_1 \mathrm{d} b_1 \int_{0}^{\infty} b_2 \mathrm{d}b_2  \tilde{V}_{g, 2}(b_1, b_2)  Z_{\mathrm{SJT}}^{\mathrm{tr}}(\beta_1, b_1) Z_{\mathrm{SJT}}^{\mathrm{tr}}(\beta_2, b_2).
\end{equation}
Above the genus-counting topological factor $e^{-2g S_0}$ accounts for the contribution of the Einstein-Hilbert action, which weights the various terms in the genus expansion.  To compute the volumes of the moduli space of super Riemann surfaces, one could use the analogous of the recurrence relations introduced by Mirzakhani for the bosonic case \cite{Mirzakhani:2006fta,Mirzakhani:2006eta} and generalized to the superspace in \cite{Stanford:2019vob}. However, for the upcoming discussion, it will not be necessary to have the explicit expressions of the volumes. We will only need to know that these volumes are polynomials in $b_i^2$ of degree $g-1$ for $g>0$ \cite{Norbury:2020vyi}. Thus we formally write
\begin{equation}\label{general_form}
\tilde{V}_{g, 2}(b_1, b_2)= \sum_{d_1, d_2 =0}^{d_1 + d_2 = g-1} \frac{\tilde{V}_{g, n}^{d_1, d_2}}{4^{d_1} 4^{d_2} d_1 ! d_2 !} b_1^{2d_1} b_2^{2d_2}.
\end{equation}
The unknown coefficients $\tilde{V}_{g, n}^{d_1, d_2}$ could be computed explicitly exploiting the results in \cite{Norbury:2020vyi}. Plugging the polynomial \eqref{general_form} into \eqref{wormhole}, we  obtain the following representation for the  genus-$g$ contribution to the wormhole amplitude:
\begin{equation}
Z_g^{\mathrm{SJT}}(\beta_1, \beta_2) = \frac{e^{-2g S_0}}{2 \pi \sqrt{\beta_1 \beta_2}} \sum_{d_1, d_2 =0}^{d_1 + d_2 = g-1} \frac{\tilde{V}_{g, n}^{d_1, d_2}}{4^{d_1} 4^{d_2} d_1 ! d_2 !}\int_{0}^{\infty} b_1 \mathrm{d} b_1 \int_{0}^{\infty} b_2 \mathrm{d}b_2 \   b_1^{2d_1} b_2^{2d_2} \   e^{-\frac{b_1^2}{4\beta_1}}  e^{-\frac{b_2^2}{4\beta_2}}
\end{equation}
By integrating over $b_1$ and $b_2$, and subsequently analytically continuing the expression to complex values of $\beta_{1,2}$, i.e., $\beta_{1,2}=\beta\pm i t$, we obtain
\begin{equation}
\label{ciro}
Z_g^{\mathrm{SJT}}(\beta_1-i t, \beta_2+i t)=\frac{2}{\pi} e^{-2g S_0}   \sum_{d_1, d_2 =0}^{d_1 + d_2 = g-1} \tilde{V}_{g, 2}^{d_1, d_2} \left(\beta+it\right)^{d_1 + \frac12} \left(\beta-it\right)^{d_2 + \frac12}.
\end{equation}
The leading behavior of \eqref{ciro} in the limit $t \rightarrow +\infty$ only comes from the terms with $d_1 + d_2 = g - 1$. Thus the leading contribution to $Z_g^{\mathrm{SJT}}$ takes the form:
\begin{equation}\label{t_infty}
Z_g^{\mathrm{SJT}}(\beta + it, \beta - it) \underset{t\rightarrow +\infty}{\approx} C_g e^{-2g S_0} t^g
\end{equation}
for some unknown constant coefficient $C_g$. If we now switch to the rescaled time variable $\tau= e^{-S_0}t$ and we perform the limit $S_{0}\rightarrow \infty$, we get for $g>0$
\begin{equation}\label{tau}
\lim_{S_0 \to \infty} e^{-S_0} Z_g^{\mathrm{SJT}}(\beta + it, \beta - it) = \lim_{S_0 \to \infty} C_g e^{-(g+1) S_0} \tau^g =0.
\end{equation}
Clearly, all the subleading terms that we have omitted in \eqref{t_infty} will also become negligible in this regime.

This proves that, in the context of SJT, all contributions to the SFF arising from surfaces of genus $g>0$ vanish in the $\tau$-scaling limit. The only exception to this scenario is the genus-zero contribution, i.e. the cylinder that is built by directly gluing two trumpets together. Its expression is given as 
\begin{equation}\label{double_tr}
Z_{g=0}^{\mathrm{SJT}}(\beta_1, \beta_2)= c\frac{\sqrt{\beta ^2+t^2}}{2 \pi  \beta }\underset{t\rightarrow +\infty}{\approx} c\frac{t}{2 \pi \beta}.
\end{equation}
Here, following the conventions in \cite{Stanford:2019vob}, we have incorporated a factor $c$ that can take on different values depending on whether we sum over spin structures and orientation reversal. In our specific case, where we sum over one but not the other and thus we set 
$c=2.$
Consequently, after substituting $t=\tau e^{S_{0}}$ and multiplying by a scaling factor of  $e^{-S_{0}}$  as indicated in  \eqref{tau}, we arrive at a finite contribution of $\frac{\tau}{\pi \beta}$  from the double-trumpet topology in the $\tau$-scaling limit. This remains the only term that survives in the genus expansion \eqref{series}.

It is worth highlighting some key differences between the current analysis and the corresponding calculations in JT gravity. In the latter, the volumes of the moduli space of Riemann surfaces with two boundaries are polynomials of degree  $3g-1$ with respect to the variables $b_1^2$ and $b_2^2$. If one were to extrapolate the current analysis to the JT gravity case,  one might naively anticipate a leading term scaling as 
$t^{3 g}$ for large $t$. However, as detailed in \cite{Blommaert:2022lbh, Weber:2022sov}, a remarkable series of highly nontrivial cancellations occur in the large time expansion of the sum:
\begin{equation}
  \sum_{d_1, d_2 =0}^{d_1 + d_2 = 3g-1} \tilde{V}_{g, 2}^{d_1, d_2} \left(\beta+it\right)^{d_1 + \frac12} \left(\beta-it\right)^{d_2 + \frac12},
\end{equation}
This effectively reduces the maximal power of  $t$ from $t^{3g}$  to $t^{2g+1}$. Consequently, the dominant term for JT gravity at genus $g$ is:
\begin{equation}
e^{-S_0} Z_g^{\mathrm{JT}}(\beta + it, \beta - it) \approx \hat{C}_g e^{-(2g+1)S_0} t^{2g+1} = \hat{C}_g \tau^{2g + 1},
\end{equation}
This term survives the $\tau$-scaling limit for each genus $g>0$, as substantiated in \cite{Blommaert:2022lbh, Weber:2022sov}.

\subsection{The nonperturbative structure}\label{Bessel}
In the previous section, we observed that, with the exception of the linear term, that causes the ramp growth in the spectral from factor, all perturbative terms in the genus expansion of the SSF for SJT gravity vanish in the $\tau$-scaling limit. However, as anticipated in \eqref{series}, this fact does not preclude the possibility of nonperturbative contributions appearing in the rescaled time $\tau$. Indeed, if the genus-zero contribution to the $\tau$-scaled spectral form factor is meaningful on its own, we expect some nonperturbative term to be responsibile of compensating the linear growth of the ramp and generating a late-time plateau.

In this section, we start looking into the nonperturbative structure of the SFF, focusing firstly on the low-energy regime of SJT gravity, where a description in terms of the more manageable Bessel model is accessible. In the same way that the pure Airy case serves as a prototype for the Saad–Shenker–Stanford description of JT gravity \cite{Saad:2019lba}, this Bessel
case should in fact play a similar role for SJT gravity.
The disk-level spectral density for SJT gravity, obtained by perfoming the inverse Laplace transform of the disk result in 	\eqref{basic}, is in fact given by
\begin{equation}\label{disk}
\rho_{0} ^{(\mathrm{SJT})}(E)= \sqrt{2} \ \frac{\cosh(2 \pi \sqrt{E}) }{\pi \sqrt{E} }
\end{equation}
which reduces indeed to the Bessel spectral density as $E \rightarrow 0$.

In order to provide a sensible nonperturbative completion of the SFF, sensitive to the underlying discreteness of the energy eigenvalues, we will exploit the matrix model description of JT supergravity, which was classified by Stanford and Witten in \cite{Stanford:2019vob} by generalizing the relationship between JT gravity and double-scaled random matrix theory \cite{Saad:2019lba}. \footnote{In this work, we will only consider SJT gravity without time reversal,  meaning that we do not have to gauge the time reversal symmetry in the bulk.} It was stated that, when fermions are introduced to the boundary theory, it is necessary to sum appropriately over the spin structures of orientable surfaces in the bulk. There are two ways to accomplish this while obeying the topological field theories' general requirements \cite{Stanford:2019vob}: one can weight the different spin structures with 1 or with $(-1)^{\zeta}$, where $\zeta$ is the Atiyah-Singer index.
These two prescriptions define two distinct gravitational theories and consequently different dual random matrix theories, which are distinguished by the properties of their supercharge $Q$. Weighting with 1 in the bulk corresponds to a theory with a random supercharge sampled from a Gaussian Unitary Ensemble (GUE), instead weighting with $(-1)^{\zeta}$ induces a supercharge of the form	$ Q =\begin{pmatrix}
0 & C^{\dagger} \\
C & 0
\end{pmatrix}$
where $C$ is a $L\times L$ complex square matrix transforming as a bifoundamental of  $U(L)\times U(L)$ and obeying the Altland-Zirnbauer statistic with $\alpha =1 $ and $ \beta = 2$. In this work, we will focus on this last case, which in the literature is also known as \textit{chiral Gaussian unitary ensemble}. From now on, we will adopt the standard notation $e^{-S_{0}}\equiv \hbar$ used in the matrix model context. 

In the following, we will be interested in the statistics governing the correlation between close eigenvalues in the aforementioned \textit{chiral Gaussian unitary ensemble}. In this context, it has been recently observed \cite{Akemann:2021uhf} that, as long as the eigenvalues are in the bulk of the spectral support, the repulsion between them is governed by the universal sine kernel, already introduced in \eqref{sine}. On the other hand, as we move towards the hard edge of the spectrum at $E=0$, the eigenvalue statistics for this model is controlled by the Bessel kernel, which we derive in detail in \eqref{CD_kernel2}. Moreover, in \cite{Akemann:2021uhf} it was also shown that, as the level spacing goes to zero, the Bessel kernel becomes well approximated by the sine kernel. In \ref{matrix} we review and rephrase the derivation by applying it specifically to the tau-scaling limit, where this uncontrolled approximation becomes exact.

As a conclusion, by virtue of the same steps that led to \eqref{SFF} through the Fourier transform of the sine kernel,  we claim that the $\tau$-scaled spectral form factor for SJT gravity is given by \footnote{From now on, we will simply call SJT gravity the theory corresponding to $(1, 2)$ Altland-Zirnbauer without time reversal. }:
\begin{equation}\label{genus-zero2}
\mathrm{SFF}_{\mathrm{SJT}}= \int_{0}^{\infty} \mathrm{d} E e^{-2 \beta E} \min\left(\frac{\cosh\left(2\pi \sqrt{E}\right)}{\pi \sqrt{2E}}, \frac{\tau}{2\pi}\right)
\end{equation} 
As a first observation, we remark that  we have not been able to perform the integral over $E$ in \eqref{genus-zero2} in closed analytic form, since this would imply finding the solutions $E^{*}$ to the trascendental equation:
\begin{equation}\label{trascendental}
\frac{\cosh\left(2\pi \sqrt{E^{*}}\right)}{\pi \sqrt{2E^{*}}}=\frac{\tau}{2\pi}
\end{equation} 
Although efficient results can be easily obtained numerically, our focus is to derive a characterization for the nonperturbative contributions hidden in the above expression and their possible geometrical interpretation. To this aim, as a first step, we can see what happens in the Bessel model, where the solution to the low energy limit of \eqref{trascendental} is readily found to be $E^{*}=2/\tau^2$.

\begin{figure}
	\centering
	\begin{tikzpicture}
	\node (image) at (0,0) {
		\includegraphics[width=0.8\linewidth]{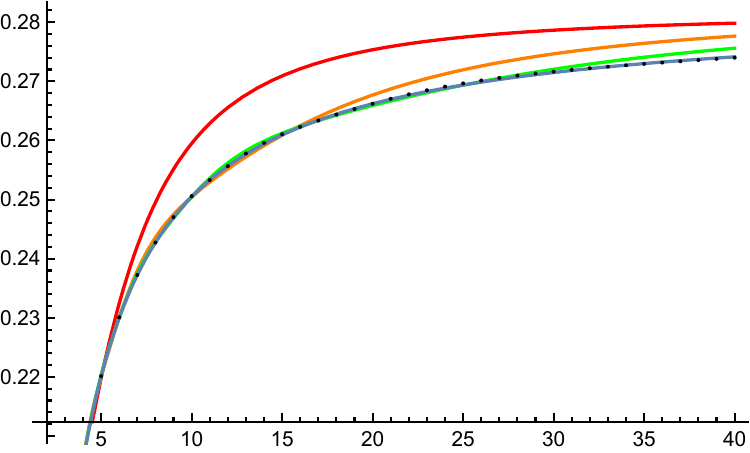}
	};
	\node at (0,-4) {$\tau$};
	\node at (-6,4) {$\mathrm{SFF}_{\mathrm{Bessel}}(\tau)$};
	\end{tikzpicture}
	\caption{The blue line corresponds to the result for SFF obtained by approximating the Bessel kernel with the sine kernel, i.e. formula \eqref{bessel_result}. The other lines correspond to numerical evaluations of the SFF in \eqref{two-point2} using \eqref{CD_kernel2} for different values of $\hbar$. In particular: red line corresponds to $\hbar=\frac{1}{10}$, orange line to $\hbar=\frac{1}{30}$, green line to $\hbar=\frac{1}{50}$. Finally, dark dots correspond to $\hbar=\frac{1}{100}$ and are almost perfectly overlapped with the analytic result, valid as $\hbar \rightarrow 0$. In this graph, $\beta=1$.}
	\label{fig:spectralformfactor}
\end{figure}
In this case, the integral over $E$ is very easy to perform
\begin{equation}\label{bessel_result}
\begin{split}
\mathrm{SFF}_{\mathrm{Bessel}}(\beta,\tau)&= \frac{\tau}{2 \pi}\int_{0}^{\frac{2}{\tau^2}} \mathrm{d} E e^{-2 \beta E} +\frac{1}{\pi}\int_{\frac{1}{\tau^2}}^{+\infty} \frac{\mathrm{d} E}{\sqrt{2E}} e^{-2 \beta E} \\
&=\frac{\tau }{4 \pi  \beta }-\frac{\tau }{4 \pi  \beta } e^{-\frac{4 \beta }{\tau ^2}}+\frac{\text{erfc}\left(\frac{2 \sqrt{\beta }}{\tau }\right)}{2 \sqrt{\pi } \sqrt{\beta }}
\end{split}\end{equation}
We note something remarkable: the only perturbative term in $\tau$ appearing in \eqref{bessel_result} is the linear one. This result is consistent with what we have found in Section \ref{bulk} from the bulk point of view, where we proved that for SJT gravity all higher orders in $\tau$ vanish identically in the $\hbar \rightarrow 0$ limit. If this property holds for SJT, it was in fact expected to hold also for its low energy limit, as we confirm here. Furthermore, the coefficient of the linear term in \eqref{bessel_result} exactly agrees with the one computed in \eqref{double_tr} from the gravitational perspective \footnote{Actually, there is a mismatch factor of $1/4$ between the linear term in \eqref{bessel_result} and the double-trumpet, that is due to the relation $Z_{\mathrm{MM}}^{\mathrm{tr}}(\beta, b)=\frac12 Z_{\mathrm{bulk}}^{\mathrm{tr}}(\beta, b)$ between the matrix model and the bulk normalizations of the trumpet partition function.}.

The novelty is that the matrix model description has provided us with a nonperturbative completion of this bulk picture, represented by the last two terms in \eqref{bessel_result}. As we can appreciate in Figure \ref{fig:spectralformfactor}, these nonperturbative terms determine a late time plateau. In Figure \ref{fig:spectralformfactor}, we have also represented some numerical evaluations of the SFF \eqref{two-point2}, by inserting the exact expression \eqref{CD_kernel2} for the Bessel kernel for different values of $\hbar$: we notice that, as $\hbar$ becomes smaller, the numerical lines tend to the analytic result \eqref{bessel_result}, as we expected from our discussion above.

A final comment is now in order. While for JT gravity and the Airy model the $\tau$-scaling limit was responsable for a perturbative plateau \eqref{perturbative_plateau} \eqref{JT}, in this case the power series in $\tau$ is completely annihilated, except for the linear term, and a plateau is produced from purely nonperturbative effects. On general ground, one would expect that nonperturbative contributions to an observable arise from the completion of "standard" perturbation theories, that are usually asymptotic in quantum systems. The peculiar behavior of \eqref{bessel_result} makes us wonder whether there is still some kind of mechanism through which we can make the nonperturbative part of the result to resurge, so that to find a connection to a genus -like expansion. A particular form of resurgence theory will enable us to realize this program, providing us with a analytic transeries approximation of the SJT $\tau$-scaled spectral form factor \eqref{genus-zero2}.

\section{Resurgence for the $\tau$-scaled SFF}
\subsection{A new contour for $P_g^{(\rho)}(\beta)$ and the Bessel model}\label{contour}

As anticipated in \ref{tau_scaling}, it was proven in \cite{Saad:2022kfe} that the general coefficient \eqref{Pn} of $\tau^{2g+1}$ generates, upon summation over $g$, the formula \eqref{summation} for the spectral form factor in the $\tau$-scaling limit. The demonstration relies on the fact that the spectral density $\rho_{0}(E)$ is a monotonic function of the energy, i.e. on the fact that there is a unique solution $E_{*}$ that satisfies $\rho(E_{*})=\frac{\tau}{2 \pi}$. 

\begin{figure}\label{new_contour}
	\centering
\begin{tikzpicture}[scale=2, thick]
\draw[fill] (0, 0) circle (.7pt) node[shift=(225:0.4)]{O};
\draw[fill] (\endfirstcut, 0) circle (.7pt) node[shift=(-100:0.5)]{\small $E_1^*$} ;
\draw[fill] (\beginsecondcut, 0) circle (.7pt) node[shift=(-80:0.5)]{\small $E_2^*$} ;
\draw[cutstyle, thick] (\endfirstcut, 0) -- (\beginsecondcut, 0);
\draw[\contourcolor, ultra thick, arrow data={0.4}{latex}, arrow data={0.9}{latex}]
(\graphwidth,-\contourgap) -- (0,-\contourgap) arc (270:90:\contourgap) -- (\graphwidth,\contourgap);
\node at (2.7,0.3) {\large{\textcolor{red}{$\gamma$}}};

\drawaxes{$\operatorname{Re}\,E$}{$\operatorname{Im}\,E$}

\end{tikzpicture}
\caption{The new contour $\gamma$ used in \eqref{Pn2} is displayed. In the case of SJT, where after summation a branch cut emerges between $E_1^*(\tau)$  and $E_2^*(\tau)$, $\gamma$ can be deformed into a new contour surrounding the branch cut.}
\end{figure}
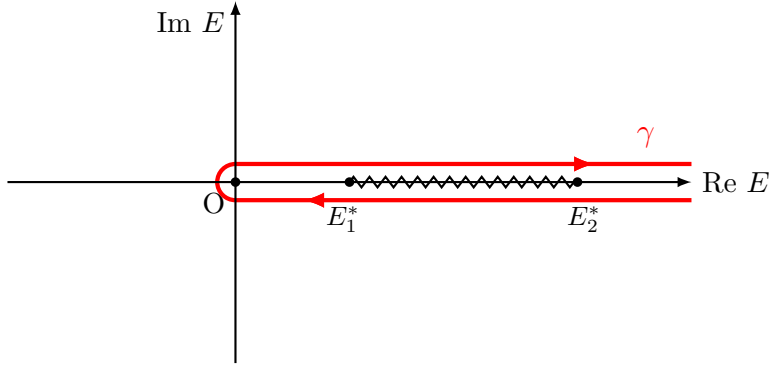

In this Section, we present an extension of \eqref{Pn} that allows to recover \eqref{summation} for more general spectral densities, such as the one relevant for SJT gravity: in this last case, $\rho_{\mathrm{SJT}}$ is not monotonic and \eqref{trascendental} has two non-trivial solutions $E_1^*(\tau)$  and $E_2^*(\tau)$. We propose then the following generalization of \eqref{Pn}:
\begin{equation}\label{Pn2}
P_g^{(\rho)}(\beta)=- \frac{1}{g (2g +1)(2\pi)^{2g+1}}\int_{\gamma} \frac{\mathrm{d}E}{2 \pi i} \frac{e^{-2 \beta E}}{\rho_0(E)^{2g}}
\end{equation}   
where $\gamma$ is the contour diplayed in Figure \ref{new_contour}, which encircles the origin and wraps the whole positive real axis\footnote{Mirroring the derivation of \cite{Saad:2022kfe}, one can in fact perform the summation over $g$ inside \eqref{Pn2} and then deform $\gamma$ into a contour surrounding the emerging branch cut between $E_1^*(\tau)$  and $E_2^*(\tau)$, the two solutions of the trascendental equation \eqref{trascendental}.}.

As a first step, we notice that by simply sobstituting the spectral density of SJT gravity into \eqref{Pn2} it is immediate to realize that the integral vanishes. This occurrence arises due to the integrand being an entire function, allowing the contour $\gamma$ to be closed without encircling any singularity along the real axis. Consequently, the integral evaluates to zero as dictated by Cauchy's theorem.
This observation indeed aligns with the findings of Section \ref{bulk}, where a bulk calculation showed that each higher genus perturbative contribution to the SFF inevitably vanishes in the $\tau$-scaling limit: the coefficients \eqref{Pn2} correctly encode the perturbative data of the bulk calculations but the information appears somewhat trivial. However, if we insist in assuming the definition \eqref{Pn2} as the geometrical input for the perturbative series, we can still gain nontrivial information out of it. For instance, we can reconstruct from it the tower of nonperturbative contributions: resurgence theory suggests to extract some asymptotic series from the SSF constructed in terms of these coefficients. To do this, we employ a slick trick: we deform the parameter $g$, initially constrained to be a natural number, to a real value by introducing a small parameter $\epsilon$, such that $g \to g + \epsilon$. In the literature, this strategy is known as \textit{Cheshire cat} resurgence \cite{Kozcaz:2016wvy, Dorigoni:2017smz,Fujimori:2022qij} \footnote{However this is the first instance, at least in our knowledge, that \textit{Cheshire cat} resurgence is used in the context of JT gravity.}: although the perturbative series truncates when we send the deformation parameter $\epsilon$ to zero, we can still reconstruct non-perturbative physics out of the perturbative data using this method.

Moreover, with this deformation, the implementation of the contour $\gamma$ becomes very natural. This is a direct consequence of introducing in \eqref{Pn2} a branch cut along the positive real axis, which acts as an obstruction to closing the contour and obtaining a vanishing result.

Let us see what happens concretely in the case of the Bessel model \footnote{To simplify numerical factors in the calculation, we find it convenient to renormalize the spectral density as $\rho_{0}(E)=\frac{1}{2 \pi \sqrt{E}}$.}. By sending $g\rightarrow g+\epsilon$ in \eqref{Pn2}, we get the following form for the coefficient $P_{g+\epsilon}^{(\rho_{\mathrm{Bessel}})}(\beta)$:
\begin{equation}
P_{g+\epsilon}^{(\rho_{\mathrm{Bessel}})}(\beta)=- \frac{1}{2\pi (g+\epsilon) (2(g+\epsilon) +1)}\int_{\gamma} \frac{\mathrm{d}E}{2 \pi i} e^{-2 \beta E} E^{g + \epsilon}.
\end{equation}
The integral can be performed by taking the discontinuity across the real axis and yields
\begin{equation}\label{coefficient}
P_{g+\epsilon}^{(\rho_{\mathrm{Bessel}})}(\beta)=- \frac{1}{2\pi^2 (g+\epsilon) (2(g+\epsilon) +1)} e^{i \pi \epsilon} \sin{(\pi \epsilon)} \frac{\Gamma(g+1+ \epsilon)}{(2 \beta)^{g+1+ \epsilon}}
\end{equation}
Crucially the coefficients of the series no longer vanish and exhibit the appropriate asymptotic behavior for large genus, growing like $g!$, which is indicative of an asymptotic series. We now proceed to perform a Borel transform of the series in $\tau$ and we get: \footnote{The usual prescription would be $\mathcal{B}\left[\sum_g a_g z^g\right](\xi)=\sum_{g}\frac{a_g}{g!}\xi^g$ but in our case we find it convenient to define it as $\mathcal{B}\left[\sum_g a_g z^{g+\epsilon}\right](\xi)=\sum_{g}\frac{a_g}{\Gamma (g+1+\epsilon)}\xi^{g+\epsilon}$ to explicitly remove the $\Gamma(g+1+\epsilon)$ factor.}
\begin{equation}
\begin{split}
\mathcal{B}\left[\sum_{g=1}^{\infty} P_{g+\epsilon}^{(\rho_{\mathrm{Bessel}})}(\beta) \ \tau^{2g+2\epsilon+1}\right](\xi)&=\frac{e^{i \pi \epsilon } \sin{(\pi \epsilon)}}{ 2 \pi^2} \ \frac{\tau}{2 \beta}  \sum_{g=1}^{\infty}  \frac{\xi^{g+\epsilon }}{(g+\epsilon) (2(g+\epsilon) +1)} \\
&=C_{\epsilon} \ \frac{\tau}{2 \beta}  \xi^{1+ \epsilon} \left( \Phi (\xi ,1,\epsilon +1)- \Phi \left(\xi ,1,\epsilon +\frac{3}{2}\right) \right)
\end{split}
\end{equation}
where we defined $C_{\epsilon}=\frac{e^{i \pi \epsilon } \sin{(\pi \epsilon)}}{ 2 \pi^2}$ and we denoted with $\Phi(\xi,s,a)= \sum_{k=0}^{\infty} \frac{\xi^k}{(k+a)^s}$ the Lerch Trascendent. A crucial property of this function is that it has a branch cut in the complex $\xi$-plane from $\xi=1$ to $\xi = \infty$. Because of that, we cannot naively perform an inverse Borel transform without running into an ambiguity, since the real axis represents now a Stokes line in the Borel plane. Nonetheless, according to Resurgence Theory, we can deal with this issue through the directional Laplace Transform $\mathcal{S}_{\vartheta}\left[\mathcal{B}(\xi)\right]$ \footnote{It is defined as $\mathcal{S}_{\vartheta}\left[\mathcal{B}(\xi)\right](x)=\frac{1}{x} \int_{0}^{e^{i\vartheta \infty}} e^{-\xi/x} \mathcal{B}(\xi) \mathrm{d}\xi$  }: in particular, the nonperturbative contribution we are looking for will be given by the difference between the directional Laplace Transform performed slightly above and slightly below the Stokes line at $\vartheta=0$. Writing the integrals explicitly, we need to evaluate
\begin{equation}\label{semi}
\begin{split}
\left(\mathcal{S}_{0^{+}}-\mathcal{S}_{0^{-}}\right)\left[\mathcal{B}(\xi)\right]&= \frac{2\beta}{\tau^2}\int_1^{\infty} \mathrm{d} \xi\, e^{-\frac{2 \beta}{\tau^2}\xi} \left[\mathcal{B}(\xi+i\delta)- \mathcal{B}(\xi - i \delta)\right] \\
&= \frac{2 \beta}{\tau^2} e^{-\frac{2 \beta}{\tau^2}}\int_0^{\infty} \mathrm{d} \xi\, e^{-\frac{2 \beta}{\tau^2}\xi} \mathrm{Disc}\left(\mathcal{B}\right) (\xi+1)
\end{split}
\end{equation}
where in the second step we shifted the integration variable.
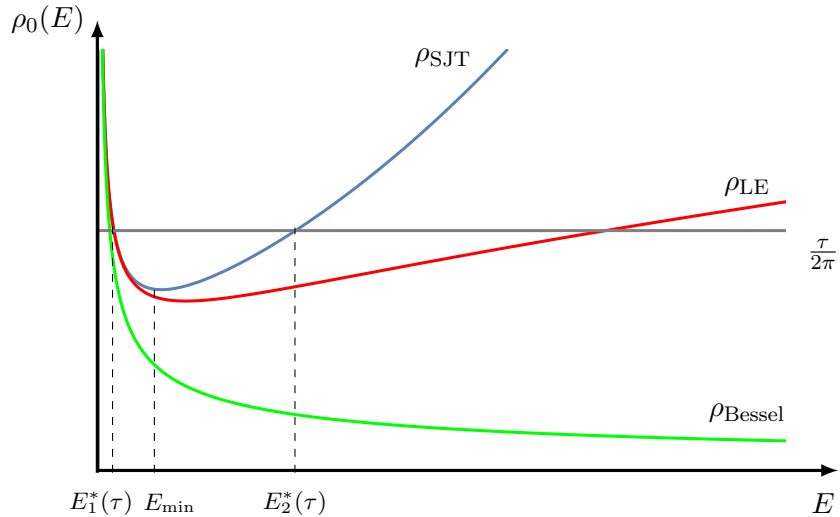
\begin{figure}
	\centering
	\begin{tikzpicture}
	\node (image) at (0,0) {
		\includegraphics[width=0.6\textwidth]{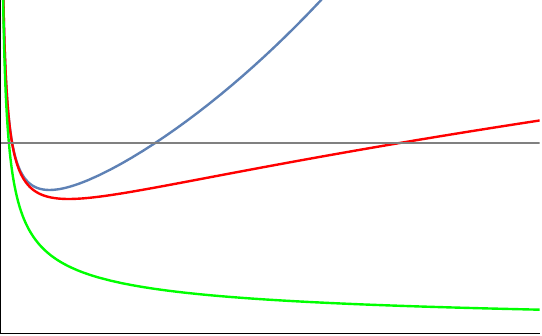}
	};
	\node at (5,0.1) {$\frac{\tau}{2 \pi}$};
	\node at (-5.2,3.2) {$\rho_{0}(E)$};
	\node at (5,-3.2) {$E$};
	\node at (4,-2) {$\rho_{\mathrm{Bessel}}$};
	\node at (0,2.7) {$\rho_{\mathrm{SJT}}$};
	\node at (4,1) {$\rho_{\mathrm{LE}}$};
	\draw[-latex,black, very thick] (-4.55,-2.77) -- (5.2,-2.77);
	\draw[-latex,black, very thick] (-4.55,-2.8) -- (-4.55,3.2);
	\draw[dashed] (-1.95,-2.8) -- (-1.95,0.45);
	\draw[dashed] (-4.35,-2.8) -- (-4.35,0.49);
	\draw[dashed] (-3.8,-2.8) -- (-3.8,-0.3);
	\node at (-4.5,-3.2) {\footnotesize{$E_{1}^{*}(\tau)$}};
	\node at (-1.95,-3.2) {\footnotesize{$E_{2}^{*}(\tau)$}};
	\node at (-3.6,-3.2) {\footnotesize{$E_{\mathrm{min}}$}};
	\end{tikzpicture}
	\caption{Plot of the spectral density $\rho_{0}(E)$ as a function of the energy: the blue line represents SJT gravity, the green represents the Bessel model, while the red line correponds to low energy expansion of SJT, capturing the first order correction to the Bessel model and the object of \ref{simplified}. The horizontal line, corresponding to $\tau/2 \pi$, determines the intersections $E_{1}^{*}(\tau)$ and $E_{2}^{*}(\tau)$ with $\rho_{\mathrm{SJT}}$.}
	\label{fig:graph2}
\end{figure}
The final ingredient we need is the jump of $\mathcal{B}(\xi)$ across the branch cut. The latter can be evaluated using the following expressions for the discontinuity of the Lerch Phi functions, which hold true when $\Re(\xi)>1$ and $\Im(\xi)=0$:
\begin{equation}\begin{split}
&\Phi(\xi- i\delta,s,a)=\Phi(\xi,s,a)   \\
&\Phi(\xi+ i \delta,s,a)= \Phi(\xi,s,a) + \frac{2 \pi i \xi^{-a} }{\Gamma(s)} \log^{s-1}(\xi)
\end{split}\end{equation}
Using this result, the discontinuity of the function $\mathcal{B}(\xi)$ across its branch cut is given by $\mathrm{Disc}(\mathcal{B}(\xi)) = 2\pi i (1 - \xi^{-\frac{1}{2}})$. Here, we choose to consider the cut of the function $\xi^{1+\epsilon}$ from $\xi=0$ to $\xi=\infty$ along the negative axis.
Performing the integral \eqref{semi}, we obtain
\begin{equation}\begin{split}\label{mike}
\left(\mathcal{S}_{0^{+}}-\mathcal{S}_{0^{-}}\right)\left[\mathcal{B}(\xi)\right]&= C_{\epsilon} \ \frac{2 \pi i}{\tau} e^{-\frac{2 \beta}{\tau^2}}\int_0^{\infty} \mathrm{d} \xi\, e^{-\frac{2 \beta}{\tau^2}\xi}   \left(1 - (\xi+1)^{-\frac{1}{2}}\right) \\
&=- (1- e^{2\pi i \epsilon} ) \left(\frac{\tau}{4 \pi \beta }  e^{-\frac{2 \beta }{\tau ^2}} - \frac{1}{2^{\frac{3}{2}} \sqrt{\pi \beta}}\text{erfc}\left(\frac{\sqrt{2\beta }}{\tau }\right) \right)
\end{split}\end{equation}
We highlight two observations regarding this result. Firstly, resurgence has revealed the hidden nonperturbative contribution in \eqref{mike}, which has exactly the form we were expecting. Secondly, as $\epsilon$ approaches zero, the overall phase jump in \eqref{mike} causes the entire result to vanish. This is known as a \textit{Cheshire cat} point, that occurs when $g$ is restored to be an integer. However, it is important to note that resurgence alone does not allow us to determine the real part of the transseries parameter. Therefore, we are unable to identify the remaining real contribution as we approach this \textit{Cheshire cat} point. The best we can do is to introduce an undetermined transseries parameter, denoted as $\sigma$ \footnote{In the context of resurgence, these constants are referred to as Stokes coefficients.}, which must be determined using external input not provided by resurgence. We express this as follows:
\begin{equation}\label{bb}
\begin{split}
\left(\mathcal{S}_{0^{+}}-\mathcal{S}_{0^{-}}\right)\left[\mathcal{B}(\xi)\right]=\sigma \left(\frac{\tau}{4 \pi \beta } e^{-\frac{2 \beta }{\tau ^2}} - \frac{1}{2^{\frac{3}{2}} \sqrt{\pi \beta}}\text{erfc}\left(\frac{\sqrt{2\beta }}{\tau }\right) \right)
\end{split}
\end{equation}
Fortunately, in this case, we have the correct result \eqref{bessel_result}, which we computed in \ref{Bessel}. By carefully comparing the different normalizations chosen for the spectral density \footnote{In Section \ref{Bessel}, we considered the spectral density $\rho_{\mathrm{Bessel}}=\frac{1}{\pi \sqrt{2E}}$, whereas here we considered $\rho_{\mathrm{Bessel}}=\frac{1}{2\pi \sqrt{E}}$. This amounts to redefining the constant $e^{S_{0}}\rightarrow \frac{1}{\sqrt{2}}e^{S_{0}}$. Consequently, we need to rescale $\tau \rightarrow \sqrt{2}\tau$. Additionally, recalling that $\mathrm{SFF}=\lim_{S_{0}\rightarrow \infty} e^{-S_{0}}\left<ZZ\right>$, there is an overall relative factor of $1/\sqrt{2}$.}, we can establish $\sigma=1$.

Thus far, resurgence has enabled us to recover a result that was already available. However, in the following sections, we will apply this machinery to SJT gravity, where we lack an exact result, unlike the Bessel model. It is in this scenario that resurgence will truly demonstrate its full potential.
\subsection{Low-energy limit: a simplified model}\label{simplified}
Before delving into the application of this formalism to SJT gravity, it is valuable to examine a simplified model, specifically its low-energy expansion. In the subsequent discussion, our focus will be on the spectral density given by:
\begin{equation}\label{low}
\rho_{\mathrm{LE}}(E)=\frac{1+\alpha E}{2 \pi \sqrt{E}}
\end{equation}
By setting $\alpha=2\pi^2$, we precisely obtain the low-energy expansion of the spectral density of SJT gravity, including the first correction to the Bessel model. Figure \ref{fig:graph2} showcases the Bessel model, the full SJT spectral density, and the newly introduced $\rho_{\mathrm{LE}}(E)$. We note that \eqref{low} interpolates between a Bessel behaviour at low energy and an Airy-like behaviour at large energies. It is not the first time that a similar behaviour is investigated in the literature \cite{Johnson:2019eik}.

Before commencing our analysis, we wish to emphasize that, given the current spectral density \eqref{low}, it is indeed possible to derive a closed-form solution for the spectral form factor by explicitly evaluating the integral in \eqref{genus-zero2} with $\rho_{\mathrm{SJT}}(E)\rightarrow \rho_{\mathrm{LE}}(E)$: the computation is contained in Appendix \ref{exact}, where also the relevant expansions are presented. Conversely, we have found more interesting to describe the resurgence method in the main text, providing insights into the transeries expansion and capturing the nonperturbative aspects of the problem in a general framework. Furthermore, the approach we are about to present will be more transparent in computing the spectral form factor of SJT, where an explicit analytic expression through the aforementioned integral representation is not feasible.

With that in mind, we proceed to compute the coefficients $P_g^{(\rho_{\mathrm{LE}})}(\beta)$ associated with this spectral density by substituting \eqref{low} into our master formula \eqref{Pn2}:
\begin{equation}
P_g^{(\rho_{\mathrm{LE}})}(\beta)=- \frac{1}{g (2g +1)(2\pi)}\int_{\gamma} \frac{\mathrm{d}E}{2 \pi i} e^{-2 \beta E}\frac{E^g}{\left(1+\alpha E\right)^{2g}}
\end{equation}
Next, we utilize the following expansion:
\begin{equation}\label{binomial}
\frac{1}{\left(1+\alpha E\right)^{2g}}=\sum_{k=0}^{\infty} \frac{\Gamma (2g+k)}{\Gamma(2g)} \frac{(-1)^k \alpha^kE^{k}}{k!}
\end{equation} 
After setting $g\rightarrow g+\epsilon$, we can perform the integral for each term of the series separately, yielding:
\begin{equation}
P_{g+\epsilon}^{(\rho_{\mathrm{LE}})}(\beta)= \frac{e^{i\pi \epsilon}\sin (\pi\epsilon)}{2 \pi^2 }\sum_{k=0}^{\infty} \frac{(-1)^k \alpha^k}{k! \left(2\beta\right)^{g+\epsilon+k+1}} \ \frac{\Gamma (2g+2\epsilon+k)}{\Gamma(2g+2\epsilon)(g+\epsilon) (2g+2\epsilon +1)} \Gamma(g+\epsilon+k+1)
\end{equation}
\begin{figure}
	\centering
	\begin{tikzpicture}
	\node (image) at (0,0) {
		\includegraphics[width=0.8\linewidth]{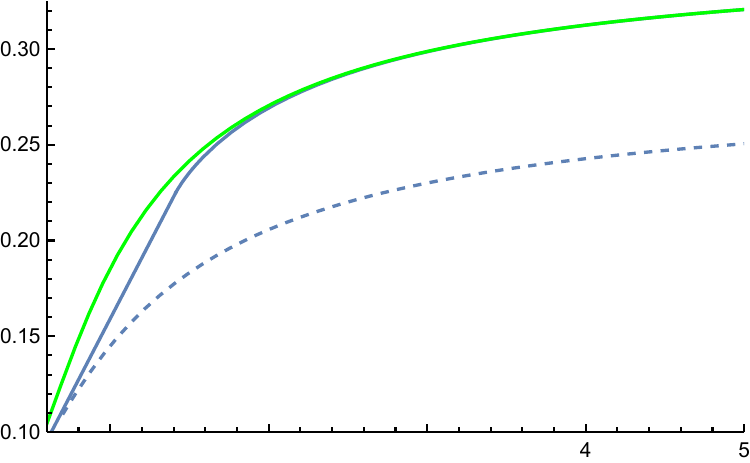}
	};
	\node at (0,-4) {$\tau$};
	\node at (0.85,-3.53) {$3$};
	\node at (-1.725,-3.53) {$2$};
	\node at (-4.27,-3.53) {$1$};
	\node at (-6,4) {$\mathrm{SFF}_{\mathrm{LE}}(\tau)$};
	\end{tikzpicture}
	\caption{We show $\mathrm{SFF}_{\mathrm{LE}}(\tau)$, for a choice of $\beta=\frac12$ and $\alpha=\frac12$. The blue line represents $\mathrm{SFF}_{\mathrm{LE}}^{\mathrm{exact}}\left(\tau\right)$, which comes from the direct evaluation of \eqref{genus-zero2} with $\rho_{\mathrm{SJT}}(E)\rightarrow \rho_{\mathrm{LE}}(E)$. We denote with $\mathrm{SFF}_{\mathrm{LE}}^{k_{\mathrm{max}}}$ the expression \eqref{bomber} where the sum over $k$ on the second line is truncated to $k_{\mathrm{max}}$. The green line corresponds to $k_{\mathrm{max}}=2$, while dashed line is the Bessel model result \eqref{bb}, or just \eqref{bomber} after setting $\alpha=0$. }
	\label{fig:berlusconi}
\end{figure}
It is noteworthy that each term of the series in $k$ continues to display the desired asymptotic behavior with respect to the genus $g$. This indicates that we can once again leverage the tools of resurgence, just as we did before.
We then perform the Borel resummation over $g$, for each term of the series over $k$ separately, i.e.
\begin{equation}\label{prodi}
\begin{split}
&\mathcal{B}\left[\sum_{g=1}^{\infty} \tau^{2g+2\epsilon+1}P_{g+\epsilon}^{(\rho_{\mathrm{LE}})}(\beta)\right](\xi)= \\
&\frac{C_{\epsilon}}{\tau}\sum_{k=0}^{\infty}  \frac{(-1)^k }{k!} \left(\frac{\alpha}{\tau^2}\right)^k \mathcal{B}\left[\sum_{g=1}^{+\infty} \left(\frac{\tau^2}{2\beta}\right)^{g+\epsilon+k+1} \ \frac{\Gamma (2g+2\epsilon+k)\Gamma(g+\epsilon+k+1)}{\Gamma(2g+2\epsilon)(g+\epsilon) (2g+2\epsilon +1)} \right](\xi)
\end{split}
\end{equation}
The Borel transform can be computed in terms of an hypergeometric $_3F_2$ function that, for each $k$, develops once again a singularity on the real axis of the $\xi$-plane. Therefore, proceeding as before, we evaluate its discontinuity across the Stokes line at $\vartheta=0$. The expression reads as follows:
\begin{equation}\label{bertinotti}
\begin{split}
&\left(\mathcal{S}_{0^{+}}-\mathcal{S}_{0^{-}}\right)\left[\mathcal{B}(\xi)\right]=\eta\sum_{k=0}^{\infty}  \frac{(-1)^k \alpha^k}{\tau^{1+2k}k!} \times \\
&\times \int_{0}^{+\infty} d \xi e^{-\frac{2\beta}{\tau^2}\xi} \mathrm{Disc}\left[\xi^{\epsilon+k}\frac{\xi  \Gamma (k+2 \epsilon +2) \, _3F_2\left(1,\frac{k}{2}+\epsilon +1,\frac{k}{2}+\epsilon +\frac{3}{2};\epsilon +2,\epsilon +\frac{5}{2};\xi \right)}{(\epsilon +1) (2 \epsilon +3) \Gamma (2 (\epsilon +1))}\right]
\end{split}
\end{equation}
where we have introduced an unfixed Stokes coefficient $\eta$ associated with the discontinuity since, arguing as before, the real part of the jump cannot be determined as we send $\epsilon \rightarrow 0$. An important criterion for fixing this ambiguity is the limit $\alpha\rightarrow 0$, which leads to the recovery of the Bessel model \eqref{bb} with only the term $k=0$ surviving in the sum. Consistency with that result allows us to set $\eta=\frac{i}{4 \pi^2}.$

Once the Stokes ambiguity has been resolved, all the information regarding the nonperturbative completion should be encoded in the remaining integral. The latter can now be evaluated by taking the limit $\epsilon\rightarrow 0$ at this stage, where the hypergeometric function simplifies drastically. We obtain:
\begin{equation}
\begin{split}
&\left(\mathcal{S}_{0^{+}}-\mathcal{S}_{0^{-}}\right)\left[\mathcal{B}(\xi)\right]=\frac{i}{4 \pi^2}\sum_{k=0}^{\infty}  \frac{(-1)^k \alpha^k}{\tau^{1+2k}k!} \int_{0}^{+\infty} d \xi e^{-\frac{2\beta}{\tau^2}\xi} \xi^{k} \mathrm{Disc}\left[f_{k}(\xi)\right]
\end{split}
\end{equation}
where we have defined $f_{k}(\xi)$ as:
\begin{equation}
\begin{split}
f_{k}(\xi)\!=\!\frac{\left(\!\left(\sqrt{\xi }+1\right)^k\!\!-\!\left(1-\sqrt{\xi }\right)^k\!\!-\!\sqrt{\xi } \left(2 (k-1) (1-\xi )^k\!+\!\left(1-\sqrt{\xi }\right)^k\!+\!\left(\sqrt{\xi }+1\right)^k\right)\!\right) \! \Gamma (k+2)}{k \left(k^2-1\right) \sqrt{\xi }(1-\xi )^{k}}
\end{split}
\end{equation}
We observe that this function exhibits a branch cut singularity at $\xi=1$ for $k=0,1$, whereas for $k\geq 2$ it only has a pole singularity at $\xi=1$. The case $k=1$ deserves separate treatment, so we will evaluate it explicitly by integrating the discontinuity of $f_{1}(\xi)$ \footnote{We observe that $f_{k}(\xi)$ is not defined for $k=1$, however $\lim_{a\rightarrow1}f_{a}(\xi)$ is well defined, so we adopt the sloppy notation $f_{1}(\xi)\equiv\lim_{a\rightarrow1}f_{a}(\xi)$ for semplicity. } across the branch cut at $\xi \in \left[1,+\infty\right.\left.\right)$:
\begin{equation}\label{cut}
\begin{split}
\left(\mathcal{S}_{0^{+}}-\mathcal{S}_{0^{-}}\right)\left[\mathcal{B}(\xi)\right]_{k=1}&=\frac{i}{4 \pi^2}\frac{(-1) \alpha}{\tau^{3}} \int_{1}^{+\infty} d \xi e^{-\frac{2\beta}{\tau^2}\xi} \xi \text{Disc}\left(\frac{2 \tanh ^{-1}\left(\sqrt{\xi }\right)}{\sqrt{\xi }}-2\right) \\
&=\frac{\alpha}{2 \pi \tau}\frac{4 \beta  e^{-\frac{2 \beta }{\tau ^2}}+\sqrt{2 \pi } \sqrt{\beta } \tau  \text{erfc}\left(\frac{\sqrt{2} \sqrt{\beta }}{\tau }\right)}{8 \beta ^2}
\end{split}
\end{equation}
\begin{figure}
	\centering
	\begin{tikzpicture}
	\node (image) at (0,0) {
		\includegraphics[width=0.8\textwidth]{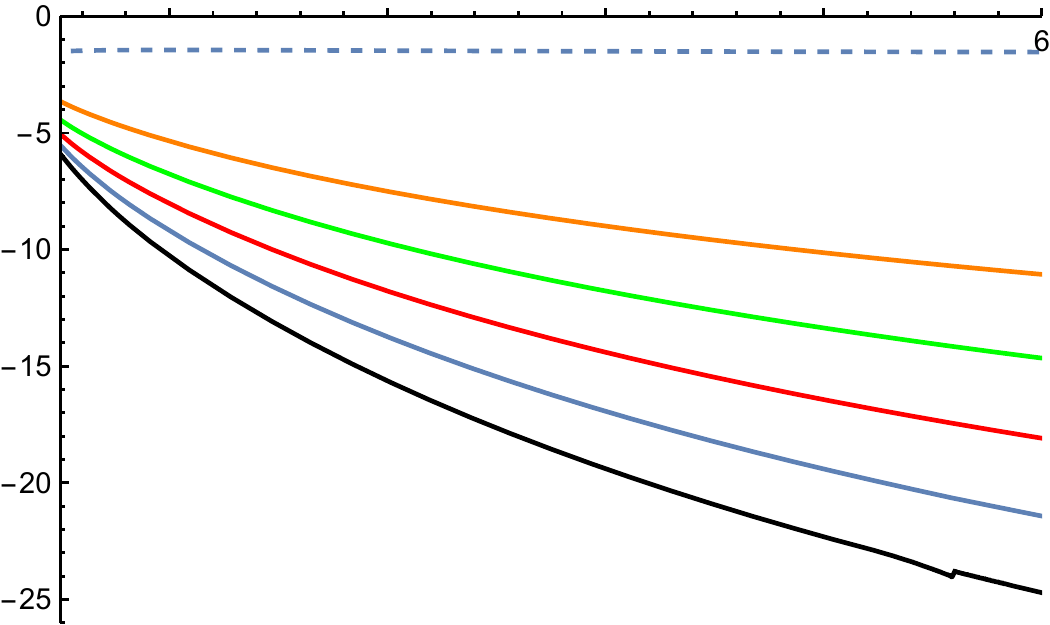}
	};
	\node at (0,4) {$\tau$};
	\node at (0.93,3.2) {$4$};
	\node at (-1.58,3.2) {$3$};
	\node at (-4.08,3.2) {$2$};
	\node at (3.45,3.2) {$5$};
	\node at (-6.9,-3.3) {$\log \left|\Delta_{\mathrm{rel}}\right|$};
	\node at (6.7,-3.3) {\footnotesize{$k_{\mathrm{max}}=5$}};
	\node at (6.7,-2.4) {\footnotesize{$k_{\mathrm{max}}=4$}};
	\node at (6.7,-1.4) {\footnotesize{$k_{\mathrm{max}}=3$}};
	\node at (6.7,-0.5) {\footnotesize{$k_{\mathrm{max}}=2$}};
	\node at (6.7,0.5) {\footnotesize{$k_{\mathrm{max}}=1$}};
	\fill [white] (4.4,-3.1) rectangle (5,-2.8);
	\draw[line width=0.52mm, black] (4.4,-2.85) -- (5,-3);
	\end{tikzpicture}
	\caption{In this Figure we show we show the increasing level of precision of our transeries \eqref{bomber}, by plotting the relative error $\Delta_{\mathrm{rel}}=\frac{\mathrm{SFF}_{\mathrm{LE}}^{\mathrm{exact}}\left(\tau\right)-\mathrm{SFF}_{\mathrm{LE}}^{\mathrm{k_{\mathrm{max}}}}\left(\tau\right)}{\mathrm{SFF}_{\mathrm{LE}}^{\mathrm{exact}}\left(\tau\right)}$  compared to the exact result. With $k_{\mathrm{max}}=1$ we mean the first line of \eqref{bomber}. Here again $\beta=\frac12$ and $\alpha=\frac12$.}
	\label{fig:pippobaudo}
\end{figure}
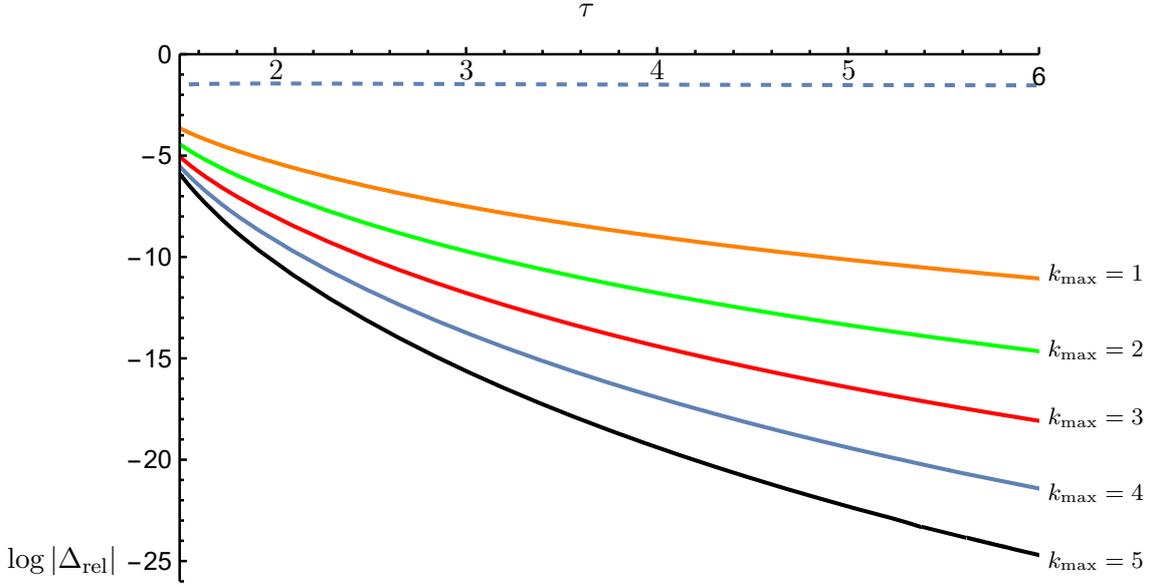
All higher orders in the expansions, for $k\geq 2$, can be obtained by evaluating the residue of $e^{-\frac{2\beta}{\tau^2}\xi} \xi^{k}f_{k}(\xi)$ at $\xi=1$. It can be expressed in terms of generalized Laguerre Polynomials as:
\begin{equation}\label{pole}
\begin{split}
\mathrm{Res}_{\xi=1}\left(e^{-\frac{2\beta}{\tau^2}\xi} \xi^{k}f_{k}(\xi)\right)=&2(k-2)!(-1)^{k-1}\exp \left(-\frac{2 \beta }{\tau ^2}\right) \sum_{n=0}^k \ (-1)^n \binom{k}{n} L_{k-1}^{\left\lfloor \frac{n}{2}\right\rfloor+1 }\left(\frac{2 \beta }{\tau ^2}\right)
\end{split}
\end{equation}
Collecting all pieces together, the result for the spectral form factor reads as follows:
\begin{equation}\label{bomber}
\begin{split}
\mathrm{SFF}_{\mathrm{LE}}(\beta,\tau,\alpha)&=\frac{\tau}{4 \pi \beta}+\theta\left(\tau-2\sqrt{\alpha}\right)\left\{\frac{\left(\alpha -\tau ^2\right) e^{-\frac{2 \beta }{\tau ^2}}}{4 \pi  \beta  \tau }+\frac{(\alpha +4 \beta ) \text{erfc}\left(\frac{\sqrt{2} \sqrt{\beta }}{\tau }\right)}{8 \sqrt{2 \pi } \beta ^{3/2}}\right. \\
&\left.-\frac{e^{-\frac{2 \beta }{\tau ^2}}}{\pi \tau} \sum_{k=2}^{\infty}  \frac{1}{k(k-1)} \left(\frac{\alpha}{\tau^2}\right)^k  \sum_{n=0}^k \ (-1)^n \binom{k}{n} L_{k-1}^{\left\lfloor \frac{n}{2}\right\rfloor+1 }\left(\frac{2 \beta }{\tau ^2}\right) \right\}
\end{split}
\end{equation}
Some comments about the result \eqref{bomber} are in order:
\begin{itemize}
\item The term in brackets encodes the nonperturbative completion extracted by resurgence. In front of this, we have also added a $\theta$ function to account for the fact that when $\frac{\tau}{2 \pi}<\rho_{\mathrm{LE}}(E_{\mathrm{min}})=\frac{\sqrt{\alpha}}{\pi}$ only the linear term of the transeries is expected to remain. In particular, the nonperturbative terms in the first line, arising from a branch cut singularity in the Borel plane, play a crucial role in determining the shape and position of the plateau in the spectral form factor. On the other hand, the terms in the second line, originating from poles in the Borel plane, accurately capture the transition from the ramp to the plateau region, since they vanish in the $\tau\rightarrow \infty$ limit.
\item In Figure \ref{fig:berlusconi}, one can appreciate the blue line as the exact result for the low energy model \eqref{low} under examination. The dashed line represents the Bessel model approximation, while the green line corresponds to the first line of \eqref{bomber} \footnote{Without considering the $\theta$ function constraint.}. Notably, even at this level of approximation, we observe a remarkable agreement with the exact result. To demonstrate the improvement achieved by including the terms in the second line of \eqref{bomber}, we provide Figure \ref{fig:pippobaudo}, which illustrates the relative error of the transseries with respect to the exact result in a logarithmic scale. The plot depicts the increasing number of corrections incorporated.
\item An important comment arises from confronting the resurgence result with the expansion of the analytical solution presented in Appendix \ref{exact}. We notice that the transeries provided by resurgence, as one can see in \eqref{bomber}, misses a term of order $\mathcal{O}\left(e^{-\frac{2 \beta \tau^2}{\alpha}}\right)$ compared to the exact result. We can understand the origin of this mismatch by noticing that the radius of convergence of the expansion \eqref{binomial} is $1/\alpha$ and the integral over $E$, overcoming the radius of convergence of the series, will in general lose nonperturbative terms in $\alpha$. The presence of this additional pieces can be traced back to the fact that $E=1/\alpha$ is exactly the point at which the spectral density \eqref{low} reverses its monotonicity. Because of that, our resurgence analysis will in general not be able to capture the second solution $E_{2}^{*}(\tau)$ at which $\rho_{\mathrm{LE}}(E)$ intersecates with $\frac{\tau}{2 \pi}$. In other words, in our approach this solution appears to be infinitely far from $E=0$, i.e. $E_{2}^{*}(\tau) \rightarrow +\infty$. However, as we can appreciate in Figure \ref{fig:pippobaudo}, the contribution arising from having the second solution $E_{2}^{*}(\tau)$ located at finite distance is suppressed on the plateu, compared to the contributions of the resurgence transeries. As we move towards the ramp instead, this nonperturbative effects in $\alpha$ become more evident. In order to capture the contributions related to the second solution one should presumably perform resurgence in the $\alpha$ parameter.
\end{itemize}

\subsection{JT supergravity}\label{SJT}
With the results of the previous section at hand, we are now ready to generalize our analysis to SJT. In this case, we must examine the following spectral density:
\begin{equation}
\rho_0(E)= \frac{\cosh\left(2\pi \sqrt{E}\right)}{2\pi \sqrt{E}}
\end{equation}
To extend the previous procedure, we need to use the power series expansion around $E=0$ for the expression:
\begin{equation}\label{expa}
\left(\frac{2\pi \sqrt{E}}{\cosh(2\pi \sqrt{E})}\right)^{2g} = (2\pi)^{2g}E^g \sum_{n=0}^{\infty} \mathcal{E}_{2n}^{(2g)}(g) \frac{(4\pi)^{2n}}{(2n)!}E^{n},
\end{equation}
where $\mathcal{E}_{2n}^{(2g)}(g)$ represents the higher-order Euler polynomials. We limit the summation to even integers due to the property $\mathcal{E}_{2n+1}^{(2g)}(g)=0$ \cite{article2}. It is worth noting that an explicit representation of these polynomials is available in terms of the Stirling numbers of the second kind \cite{article2}:
\begin{equation}\label{expa2}
\mathcal{E}_{2n}^{(2g)}(g)=  \sum_{k=0}^{2n} \binom{2n}{k} g^{2n-k} \sum_{l=0}^{k}(-1)^l \binom{2g+l-1}{l} 2^{-l} l! S(k, l) \; .
\end{equation}
Exploiting the expansion \eqref{expa}, we can compute the coefficients $P_{g+\epsilon}^{(\rho_{\mathrm{SJT}})}(\beta)$ as:
\begin{equation}\begin{split}
P_{g+\epsilon}^{(\rho_{\mathrm{SJT}})}(\beta)&=- \frac{1}{(g+\epsilon) (2g+2\epsilon +1)(2\pi)^{2} i} \sum_{n=0}^{\infty} \mathcal{E}_{2n}^{(2g+2\epsilon)}(g+\epsilon) \frac{(4\pi)^{2n}}{(2n)!} \int_{\gamma} \mathrm{d}E e^{-2 \beta E} E^{n+g+\epsilon}\\
&=- \frac{e^{i\pi \epsilon}\sin(\pi\epsilon)}{2\pi^2(g+\epsilon) (2g+2\epsilon +1) } \sum_{n=0}^{\infty} \mathcal{E}_{2n}^{(2g+2\epsilon)}(g+\epsilon) \frac{(4\pi)^{2n}}{(2n)!} \frac{\Gamma(n+g+\epsilon+1)}{(2\beta)^{n+g+\epsilon+1}} \; .
\end{split}\end{equation}
Recalling the definition \eqref{expa2}, we can bring the sums over $n,k,l$ outside and obtain the following asymptotic series:
\begin{equation}
\sum_{g=1}^{\infty} P_{g+\epsilon}^{(\rho_{\mathrm{SJT}})}(\beta)  \tau^{2g+2\epsilon +1}=  \lim_{\nu \to 0}\mathcal{D} \left(\sum_{l=0}^{k}(-2)^{-l}  S(k, l)  \mathcal{I}_{n,l} (\nu,\epsilon) \right)
\end{equation}
where we have defined the formal differential operator $\mathcal{D}$ as:
\begin{equation}
\mathcal{D}=\sum_{n=0}^{\infty} \frac{(4\pi)^{2n}}{(2n)!} \sum_{k=0}^{2n} \binom{2n}{k}   \ \frac{\text{d}^{2n-k}}{\text{d}\nu ^{2n-k}} 
\end{equation}
and the quantity $\mathcal{I}_{n,l}(\nu,\epsilon)$ as:
\begin{equation}\begin{split}
\mathcal{I}_{n,l}(\nu,\epsilon)&= -\frac{e^{i\pi \epsilon}\sin(\pi\epsilon)}{2 \pi^2 \tau^{2n+1}}\sum_{g=1}^{\infty}  \frac{\Gamma(n+g+\epsilon+1) \Gamma(2(g+\epsilon)+l)}{(g+\epsilon) (2(g+\epsilon) +1)\Gamma(2(g+\epsilon))} \ e^{(g+\epsilon) \nu} \left(\frac{\tau^2}{2 \beta}\right)^{g+\epsilon+n+1} 
\end{split}\end{equation}
and we will focus on $\mathcal{I}_{n,l}(\nu,\epsilon)$ to conduct the resurgence analysis.
Remarkably, we note the insertion of the differential operator has made $\mathcal{I}_{n,l}(\nu,\epsilon)$ look very similar to the series \eqref{prodi} that we encountered in the previous section. Because of this observation, we will manage to perform the Borel resummation in terms of the same special functions.
\begin{figure}
	\centering
	\begin{tikzpicture}
	\node (image) at (0,0) {
		\includegraphics[width=0.8\textwidth]{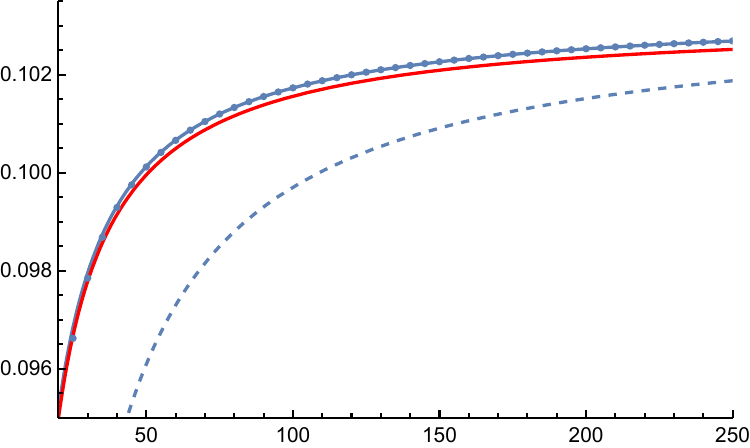}
	};
	\node at (0,-3.7) {$\tau$};
	\node at (-6,4) {$\mathrm{SFF}_{\mathrm{SJT}}(\tau)$};
	\end{tikzpicture}
	\caption{In this plot we study $\mathrm{SFF}_{\mathrm{SJT}}(\tau)$ focusing on the region near the late-time plateau, by comparing the numerical analysis with the analytical results obtained in \eqref{bomber33} and \eqref{bomber4}, for a choice of $\beta=10$. The dots represent $\mathrm{SFF}_{\mathrm{SJT}}^{\mathrm{num}}\left(\tau\right)$, which is the numerical evaluation of \eqref{genus-zero2}. We denote with $\mathrm{SFF}_{\mathrm{SJT}}^{n_{\mathrm{max}}}$ the expression \eqref{bomber33} where the sum over $n$ on the second line is truncated to $n_{\mathrm{max}}$. Red line corresponds to $n_{\mathrm{max}}=3$, while blue line to $n_{\mathrm{max}}=5$: we note that, as $n_{\mathrm{max}}$ increases, the approximation gets better. The dotted line corresponds to the late-time approximation $\mathrm{SFF}_{\mathrm{SJT}}^{\mathrm{late \ time}}(\tau)\simeq \frac{\tau}{4 \pi \beta}\left(1-e^{-\frac{2 \beta }{\tau ^2}}\right)+\frac{e^{\frac{\pi ^2}{2 \beta }}}{2 \sqrt{2 \pi \beta }} \, \text{erfc}\left(\frac{\sqrt{2 \beta }}{\tau }\right)$.}
	\label{zambrotta}
\end{figure}
Following the same logic that brought to the result \eqref{bertinotti}, we arrive at the following expression for the discontinuity of the Borel resummation of $\mathcal{I}_{n,l}(\nu,\epsilon)$:
\begin{equation}
\begin{split}
&\left(\mathcal{S}_{0^{+}}-\mathcal{S}_{0^{-}}\right)\left[\mathcal{B}(\xi)\right]=\frac{i}{4 \pi^2\tau^{2n+1}} e^{-\nu (n+1)}\times \\
&\times\int_{0}^{+\infty} d \xi e^{-\frac{2\beta}{\tau^2} e^{-\nu}\xi} \mathrm{Disc}\left[\frac{ \Gamma (l+2 \epsilon +2) \xi ^{n+\epsilon +1} \, _3F_2\left(1,\frac{l}{2}+\epsilon +1,\frac{l}{2}+\epsilon +\frac{3}{2};\epsilon +2,\epsilon +\frac{5}{2};\xi \right)}{(\epsilon +1) (2 \epsilon +3) \Gamma (2 (\epsilon +1))}\right]
\end{split}
\end{equation}
Once again, it turns out that in the cases $l=0$ and $l=1$ the function in the brackets features a branch cut, while for $l\geq 2$ the function has a pole. Performing the same steps as in \eqref{cut} and \eqref{pole} and taking the limit $\epsilon \to 0$, we are left with a purely nonperturbative term $\tilde{\mathcal{I}}_{n,l}(\nu)$ given by:
\begin{equation}
\tilde{\mathcal{I}}_{n,l}(\nu)=\begin{cases}
-\frac{1}{2 \pi }\left[ \tau\left(\frac{1}{2 \beta} \right)^{n+1} \Gamma \left(n+1,\frac{2\beta e^{- \nu }}{\tau^2}\right)-e^{-\frac{\nu }{2}} \left(\frac{1}{2\beta} \right) ^{n+\frac{1}{2}} \Gamma \left(n+\frac{1}{2},\frac{2\beta e^{- \nu }}{\tau^2}\right)\right], & l=0 \\
-\frac{1}{2\pi}e^{-\frac{\nu }{2}} \left(\frac{1}{2\beta} \right) ^{n+\frac{1}{2}} \Gamma \left(n+\frac{1}{2},\frac{2 e^{-\nu } \beta }{\tau ^2}\right), & l=1 \\
-\frac{e^{-\nu (n+1)}}{\pi\tau^{2n+1}}  (-1)^{l} (l-2)! e^{-\frac{2 \beta  e^{-\nu}}{\tau ^2}} \sum _{m=0}^l (-1)^m \binom{l}{m}  \ L_{l-1}^{\left\lfloor \frac{m}{2}\right\rfloor-l+n+1 }\left(\frac{2\beta e^{-\nu}}{\tau ^2}\right), &  l\geq 2 \\
\end{cases}
\end{equation}
Therefore, collecting all pieces together, we arrive at the following expression for the SFF of SJT gravity:
\begin{equation}\label{bomber2}
\begin{split}
\mathrm{SFF}_{\mathrm{SJT}}(\beta,\tau)&=\frac{\tau}{4 \pi \beta}+\lim_{\nu \to 0}\sum_{n=0}^{\infty} \frac{(4\pi)^{2n}}{(2n)!} \sum_{k=0}^{2n} \binom{2n}{k}   \ \frac{\text{d}^{2n-k}}{\text{d}\nu ^{2n-k}} \left(\sum_{l=0}^{k}(-2)^{-l}  S(k, l)  \tilde{\mathcal{I}}_{n,l} (\nu) \right)
\end{split}
\end{equation}
The above expression already displays the full result but the underlying non-perturbative ingredients are not apparent in this form.
In order to obtain a more explicit expression for $\mathrm{SFF}_{\mathrm{SJT}}(\beta,\tau)$, it is necessary to know how the derivative of order $2n-k$ acts on the fundamental blocks appearing in $\tilde{\mathcal{I}}_{n,l}$. In \ref{degregori} we compute their general form and below we will use it to characterize the nonperturbative structure of $\mathrm{SFF}_{\mathrm{SJT}}(\beta,\tau)$.  

\begin{figure}
	\centering
	\begin{tikzpicture}
	\node (image) at (0,0) {
		\includegraphics[width=0.8\textwidth]{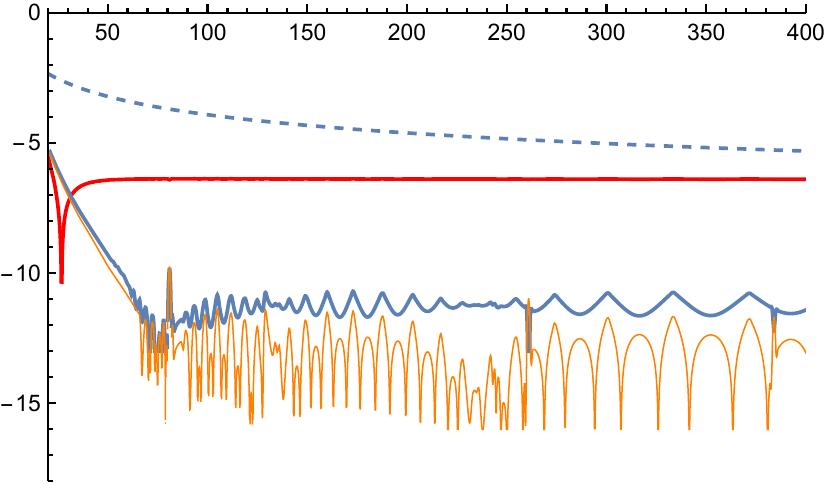}
	};
\node at (0,4) {$\tau$};
\node at (-6.5,-3) {$\log \left|\Delta_{\mathrm{rel}}\right|$};
\node at (6.6,0.9) {\footnotesize{$n_{\mathrm{max}}=3$}};
\node at (6.6,-1.1) {\footnotesize{$n_{\mathrm{max}}=5$}};
	\end{tikzpicture}
	\caption{In this Figure we show the increasing level of precision of our transeries \eqref{bomber33} for SJT, by plotting the relative error $\Delta_{\mathrm{rel}}=\frac{\mathrm{SFF}_{\mathrm{SJT}}^{\mathrm{num}}\left(\tau\right)-\mathrm{SFF}_{\mathrm{SJT}}^{\mathrm{n_{\mathrm{max}}}}\left(\tau\right)}{\mathrm{SFF}_{\mathrm{SJT}}^{\mathrm{num}}\left(\tau\right)}$. The dotted line corresponds to the relative error of $\mathrm{SFF}_{\mathrm{SJT}}^{\mathrm{late \ time}}(\tau)$, while the orange line represents the improved late-time approximation \eqref{bomber45}. Here again $\beta=10$. Spikes in the plot correspond to points where $\Delta_{\mathrm{rel}}\rightarrow0$, i.e the transeries approximation exactly coincides with the numerical result. }
	\label{balotelli}
\end{figure}

In fact, by manipulating \eqref{bomber2} as shown in \ref{degregori}, we finally arrive at the final trans-series expansion for the $\mathrm{SFF}_{\mathrm{SJT}}$:
\begin{equation}\label{bomber33}
\begin{split}
\mathrm{SFF}_{\mathrm{SJT}}(\beta,\tau)&=\frac{\tau}{4 \pi \beta}\left(1-e^{-\frac{2 \beta }{\tau ^2}}\right) +  \, \frac{1}{2 \sqrt{2\pi \beta}}\text{erfc}\left(\frac{\sqrt{2 \beta }}{\tau }\right) + \\
&+\frac{1}{\pi}\sum_{n=1}^{\infty} \frac{1}{(2n)!}\left(\frac{4\pi}{\sqrt{2\beta}}\right)^{2n}\left[\frac{\Gamma \left(n+\frac{1}{2}\right)}{2^{2n+1}\sqrt{2\beta}} \ \text{erfc}\left(\frac{\sqrt{2 \beta }}{\tau }\right)+ \frac{e^{-\frac{2 \beta }{\tau ^2}}}{ \tau}\mathcal{Z}_{n}\left(\frac{2\beta}{\tau^2}\right)\right]
\end{split}
\end{equation}
where $\mathcal{Z}_{n}\left(\frac{2\beta}{\tau^2}\right)$ is a polynomial of order $2n-2$ whose explicit form is presented in \eqref{Z}. The first few polynomials are displayed in \eqref{Z1}. In the first line of \eqref{bomber33} we recognize the contribution of the Bessel model \eqref{bb}, while the second line presents an infinite series of novel nonperturbative corrections. In Figure \ref{zambrotta} we show the remarkable agreement between the numerical evaluation of the spectral form factor and some approximations of \eqref{bomber33} with an increasing number of terms included. Moreover, in Figure \ref{balotelli} one can quantitatively appreciate the quality of the approximation by noticing how the relative error becomes smaller and smaller.

By performing the sum over $n$ for the term $\propto \text{erfc}\left(\frac{\sqrt{2 \beta }}{\tau }\right)$ in \eqref{bomber33}, one can finally rewrite the above as:
\begin{equation}\label{bomber4}
\begin{split}
\mathrm{SFF}_{\mathrm{SJT}}(\beta,\tau)&\!=\!\frac{\tau}{4 \pi \beta}\left(1-e^{-\frac{2 \beta }{\tau ^2}}\right) \!+\! \frac{e^{\frac{\pi ^2}{2 \beta }}}{2 \sqrt{2 \pi \beta }} \, \text{erfc}\left(\frac{\sqrt{2 \beta }}{\tau }\right) \!+\! \frac{e^{-\frac{2 \beta }{\tau ^2}}}{\pi \tau}\sum_{n=1}^{\infty} \frac{1}{(2n)!}\left(\frac{4\pi}{\sqrt{2\beta}}\right)^{\!2n}\!\!\!\mathcal{Z}_{n}\!\left(\frac{2\beta}{\tau^2}\right)
\end{split}
\end{equation}
In this new form of the result, we recognize the second term  as the leading contribution $\tau \rightarrow \infty$, since all other terms gets suppressed in this regime, influencing instead the transition between ramp and plateau \footnote{We point out that, as it happened for the simplified model in \eqref{bomber}, all the non-perturbative terms in \eqref{bomber4} should be multiplied by a $\theta$ function. In fact, below a certain threshold for the value of $\tau$, only the ramp is expected to remain in the expansion. }. Actually, exploiting the expression \eqref{Z2}, we can perform the sum over $n$ for the zeroth order term in $\mathcal{Z}_{n}\left(\frac{2\beta}{\tau^2}\right)$ and thus obtain an improved late-time approximation given by:
\begin{equation}\label{bomber45}
\begin{split}
\mathrm{SFF}_{\mathrm{SJT}}(\beta,\tau)&\!=\!\frac{\tau}{4 \pi \beta}\left(1-e^{-\frac{2 \beta }{\tau ^2}}\right) \!+\! \frac{e^{\frac{\pi ^2}{2 \beta }}}{2 \sqrt{2 \pi \beta }} \, \text{erfc}\left(\frac{\sqrt{2 \beta }}{\tau }\right)\! +\! \left(e^{\frac{\pi ^2}{2 \beta }}-1\right) \frac{e^{-\frac{2 \beta }{\tau ^2}}}{\pi \tau}\!+\! \mathcal{O}\left(\tau^{-3}e^{-\frac{2 \beta }{\tau ^2}}\right)
\end{split}
\end{equation}
We can appreciate the quality of this approximation by looking at Figure \eqref{balotelli}.

We conclude by remarking that also in the SJT case we expect that the resurgence approach is not able to capture the contribution from the second solution $E_{2}^{*}(\tau)$ at which $\rho_{\mathrm{SJT}}(E)$ intersecates with $\frac{\tau}{2 \pi}$: at variance with the simplified model previously discussed, here we do not have a parameter controlling the missing contributions and determining their exact behaviour is much more difficult. We did not attempt their evaluation but we are nevertheless confident that their presence should not affect significantly the formation of the plateau, as it can be appreciated from the numerical analysis.

\section{Conclusions and Outlook}\label{frodeno}

The principal aim of this paper was to explicitly examine the spectral form factor in a supersymmetric version of Jackiw-Teitelbom gravity, using the $\tau$-scaling limit proposed in \cite{Saad:2022kfe}. The main virtue of this approach, in the purely bosonic case, was to recover the expected plateau structure resumming the perturbative genus contributions to the SFF (see also \cite{Blommaert:2022lbh} for a similar computation); the emerging series in the $\tau$ variable has a finite radius of convergence and suggests a truly geometric interpretation for the late-time behaviour of SFF in JT-gravity. In the supersymmetric case we have observed that a straightforward generalization of the $\tau$-scaling limit seems to lead nowhere, because the structure of the supermoduli volumes stops the perturbative series after the first term, producing just the ramp behaviour. 
On the other hands, exploiting the relation between SJT gravity and the chiral gaussian ensemble, we argued that the SFF in the $\tau$-scaling limit can be extracted from the same integral formula governing the bosonic case, simply adapting to the SJT spectral density. We expected therefore the appearance of truly nonperturbative contribution in $\tau$. Studying the low-energy description of super JT-gravity, the Bessel model, and its first order correction we learnt that non-perturbative contributions in $\tau$ are responsible for flattening the ramp and the correct non-perturbative contributions were obtained using Cheshire cat resurgence \cite{article1, Dorigoni:2017smz}. Basically we deformed the coefficient of the perturbative expansion, encoded into the functions $P_g^{(\rho)}$,  by an analytic continuation in the genus, generating an asymptotic series that has been completed according resurgence theory. This procedure accurately reproduced the exact behaviour of our toy models and we applied the same strategy to the full super JT-gravity. Our final result is \eqref{bomber4}, expressing analytically the spectral form factor in the $\tau$-scaling limit: taking into account the non-perturbative completion  we were also able to describe the transition between the ramp and the plateau. 
Although based on some assumptions suggested by the low-energy description, our computations seem to fit perfectly with the physical expectations: on the other hand, it would be nice to check our results against different methodological approaches. The spectral form factor in super JT-gravity has been previously studied in \cite{Johnson:2020exp}, where explicit numerical computations have been performed using a non-perturbative formulation of the theory. It would be very interesting to compare, if possible, our analytical calculations with the numerical curves presented in \cite{Johnson:2020exp}. Another aspect that should be explored in the future concerns the geometrical interpretation of the non-perturbative contributions extracted from resurgence theory. In particular the coefficient of the exponentially suppressed terms in \eqref{bomber4} are polynomials in $1/\tau$, suggesting the relevance of geometries with negative Euler characteristic: this feature could be explained as the effect of disconnected surfaces, as sometimes invoked in the literature \cite{Saad:2019lba}. 
In order to better understand the analytic deformation in the genus parameter, it would be nice to obtain the formula of the coefficients $P_g^{(\rho)}$ solely using a geometrical approach: this could probably be easier in the bosonic JT case. A natural follow-up of these investigations would be the study of the spectral form factor in ${\cal N}=2$ super JT-gravity: in this theory the perturbative topological expansion in terms of Euclidean surfaces has been discussed recently by Turiaci and Witten \cite{Turiaci:2023jfa} and some nonperturbative properties have been examined \cite{Johnson:2023ofr}. It would be interesting to explore the $\tau$-scaling limit in this case.

\appendix
\section{Validity of the sine-kernel statistics for SJT gravity}\label{matrix}
In general, the Altland-Zirnbauer $(\alpha, \beta)$ matrix model is characterized by the following integral measure:
\begin{equation}
D\lambda= \prod_{i<j}\left| \lambda_i-\lambda_j \right|^\beta \prod_i \lambda_i^{\frac{\alpha -1}{2}} e^{-LV\left(\lambda_i\right)} \mathrm{d}\lambda_i   \qquad , \quad  \lambda_i > 0
\end{equation}
where $\lambda_i$ are the doubly-degenerate eigenvalues of the Hamiltonian $H=Q^{2}$.
In our case of interest, it reduces to:
\begin{equation}\label{measure}
D\lambda= \prod_{i<j}\left| \lambda_i-\lambda_j \right|^2 e^{-LV\left(\lambda_i\right)} \mathrm{d}\lambda_i   \qquad , \quad  \lambda_i > 0
\end{equation}
We note this is just the usual measure for unitary matrix models, except for the requirement of positivity of the eigenvalues. 

Actually, the infinite support of the spectral density in \eqref{disk} suggests that the suitable random matrix theory dual to SJT gravity has to be a double-scaled matrix model where we send the highest eigenvalue $\lambda_{\mathrm{max}}$ and the matrix dimension $L$ to infinity,  while keeping fixed the ratio $\frac{L}{\lambda_{\mathrm{max}}}$. As in the case of ordinary JT gravity, the potential $V(H)$ must then be tuned in such a way to obtain the desired leading order spectral density \eqref{disk}.

In particular, in \cite{Okuyama:2020qpm} it was observed that the matrix model of JT supergravity without time-reversal symmetry is equivalent to a Brezin-Gross-Witten model, once that we introduce an infinite number of couplings $t_{k}$ to the matrix model potential. This derivation was parallel to the one previously perfomed in \cite{Okuyama:2019xbv,Okuyama:2020ncd}, where it was shown that the JT gravity matrix model is a special case of the Kontsevich-Witten topological gravity, in the background where infinitely many couplings are turned on to a specific value.

The Airy and Bessel model, which are respectively the low-energy limits of JT and SJT, can be easily implemented in this formalism with a further restriction of the non-vanishing background couplings. In the following, we will briefly review the derivation of \cite{Okuyama:2020qpm} in this framework and finally perform the $\tau$-scaling limit.

We consider the free energy of the BGW matrix model $F(t_{k})$, as a function of the unspecified background couplings $t_{k}$. We then define for later use:
\begin{equation}
u:= \hbar^2 \partial_{0} F({t_k}) , \qquad \partial_k := \frac{\partial}{\partial t_k}
\end{equation}
The fundamental equation for our purpose is the so called string equation, a highly non linear differential equation for the potential $u$:
\begin{equation}\label{string_equation}
\frac{\hbar}{4} [(\partial_0\mathcal{R})^2 - 2 \mathcal{R}\partial_0^2\mathcal{R}] -2 u \mathcal{R}^2=0
\end{equation}
which originally arose from double–scaling limits of complex matrix models \cite{Morris:1990bw,Dalley:1991qg, Dalley:1992br, Klebanov:2003wg}. The quantity $\mathcal{R}$ in \eqref{string_equation} is a linear combination of Gelfand-Dickii polynomials in $u$
\begin{equation}
\mathcal{R}= \sum_{k=0}^{\infty} \hat{t}_k R_k[u] \qquad \hat{t}_k= t_k - \delta_{k, 0}
\end{equation}
As anticipated before, in order to obtain the SJT gravity theory, we have to set the couplings to the specific values $t_{0}=0,  t_k= \frac{(-1)^{k-1}}{(k-1)!}$, while the Bessel model would correspond to the trivial background $t_n=0$ for each $n \geq 0$. However, as proposed in \cite{Okuyama:2020qpm}, it is useful to leave $t_0 \neq 0$ free and fix it at last. In the latter case, the string equation \eqref{string_equation} admits an exact solution for the potential, namely:
\begin{equation}
u(x)= \frac{\hbar^2 }{8x^2}, 
\end{equation}
with $x= 1-t_0$. With a known $u(x)$, it is useful to define an auxiliary quantum mechanical system by introducing an Hamiltonian of this form:
\begin{equation}
\mathcal{H}= -\frac{\hbar^2}{4} \partial_x^2 -u(x)
\end{equation}
The operator $\mathcal{H}$ can be diagonalized and its eigenfunctions $\psi(x,E)$, called Baker-Akhiezer functions, allow us to extrapolate the fully nonperturbative physics of the model. In the case of our interest, the Bessel model, they are given by: 
\begin{equation}\label{bess_eigen}
\psi(x) = \braket{x|E} = \frac{\sqrt{x}}{\hbar} J_0\left(\frac{x\sqrt{2E}}{\hbar} \right)
\end{equation}
where $J_0(x)$ is the Bessel function of the first kind. We can now define the projector
\begin{equation}
\Pi:= \int_{0}^{1} \ket{x}\bra{x} \mathrm{d}x
\end{equation}
through which one can easily express the partition function and the connected two-point function \cite{Okuyama:2020qpm}:
\begin{equation}\label{one-two-point}
Z(\beta)= \Tr\left(e^{-\beta\mathcal{H}} \Pi \right) \qquad Z(\beta_1, \beta_2)_{c} = \Tr\left[e^{-\beta_1\mathcal{H}}\left(1- \Pi\right) e^{-\beta_2 \mathcal{H}}\Pi\right]
\end{equation}
In our discussion, we will be considering the second object which, after analytical continuation of the inverse temperatures as $\beta_{1,2}=\beta \pm i t$, will provide us with the desidered spectral form factor as a function of the time $t$. 
We are interested in studying the late-time behavior of this observable in the $\tau$ scaling limit, so we set $t=\tau/\hbar$ and we will keep only the leading term in the $\hbar \rightarrow 0$ limit. In terms of the rescaled time variable $\tau$, the two-point function in \eqref{one-two-point} reads:
\begin{equation}\label{two-point}
\begin{split}
Z\left(\beta+i\tau/\hbar, \beta - i \tau/\hbar\right)  &= \Tr\left(e^{-2\beta\mathcal{H}}\Pi\right) - \Tr\left(e^{-(\beta+i\frac{\tau}{\hbar})\mathcal{H}} \Pi e^{-(\beta - i\frac{\tau}{\hbar}) \mathcal{H}}\Pi\right)    
\end{split} 
\end{equation}
Introducing the new variables $E=\frac{E_1+E_2}{2}$, $\omega=\frac{E_2-E_1}{\hbar}$, the above is rewritten in this way:
\begin{equation}\label{two-point2}
\begin{split}
Z\left(\beta+i\tau/\hbar, \beta - i \tau/\hbar\right)  &=\int_{0}^{+\infty} \mathrm{d}E e^{-2\beta E} \rho (E)  \\
&-\hbar \int_{0}^{+\infty} \mathrm{d}E e^{-2\beta E} \int_{-\frac{2 E}{\hbar}}^{\frac{2 E}{\hbar}} \mathrm{d}\omega e^{i\omega \tau} K\left(E + \frac{\hbar \omega}{2}, E - \frac{\hbar \omega}{2}\right)^2   
\end{split} 
\end{equation}
where we denoted the spectral density and Christoffel- Darboux kernel as:
\begin{equation}\label{CD_kernel}
\begin{split}
\rho(E)&= \bra{E}\Pi \ket{E} = \int_{0}^{1} \mathrm{d}x  \ \left|\psi(x;E)\right|^2\\
K\left(E_1, E_2\right)&= \bra{E_1}\Pi \ket{E_2} = \int_{0}^{1} \mathrm{d}x  \ \psi(x;E_1)\psi(x;E_2)
\end{split}
\end{equation}
By plugging into \eqref{CD_kernel} the eigenfunctions \eqref{bess_eigen}, we obtain an explicit form for the Bessel kernel:
\begin{equation}\label{CD_kernel2}
K_{\mathrm{Bessel}}\left(E_1, E_2\right)= \frac{\sqrt{E_1} J_1\left(\frac{\sqrt{2 E_1} }{\hbar }\right) J_0\left(\frac{\sqrt{2 E_2} }{\hbar }\right)-\sqrt{E_2} J_0\left(\frac{\sqrt{2 E_1} }{\hbar }\right) J_1\left(\frac{\sqrt{2 E_2} }{\hbar }\right)}{\sqrt{2} \hbar  (E_1-E_2)}
\end{equation}
We observe that when $\hbar$ is small, the argument of the Bessel functions goes to infinity; we can thus exploit the asymptotic expansion for the Bessel functions in this limit. In particular, one has
\begin{equation}
J_1\left(x_1\right) J_0\left(x_2\right)=\frac{1}{\pi \sqrt{x_1 x_2}} \left[\sin (x_1-x_2)-\cos(x_1+x_2)+\mathcal{O}(1/x)\right]
\end{equation}
when $x_1,x_2$ are large. Therefore the Bessel kernel in this regime becomes
\begin{equation}\label{CD_kernel3}
K_{\mathrm{Bessel}}\left(E_1, E_2\right)\simeq\frac{1}{2 \pi \left(E_1 E_2\right)^{\frac14}} \left[\frac{\sin \left(\frac{\sqrt{2 E_1}}{\hbar}-\frac{\sqrt{2 E_2}}{\hbar}\right)}{\sqrt{E_1}-\sqrt{E_2}}-\frac{\cos \left(\frac{\sqrt{2 E_1}}{\hbar}+\frac{\sqrt{2 E_2}}{\hbar}\right)}{\sqrt{E_1}+\sqrt{E_2}}\right]
\end{equation}
When setting $E_{1,2}=E\pm \frac{\hbar \omega}{2}$, we immediately note the second term in \eqref{CD_kernel3} is suppressed by a power of $\hbar$ compared to the first one and therefore can be neglected in the $\hbar \rightarrow 0$ limit. When expanding the first term in the same regime, the leading contribution yields
\begin{equation}\label{Bessel_sin}
K_{\mathrm{Bessel}}\left(E + \frac{\hbar \omega}{2}, E - \frac{\hbar \omega}{2}\right)=\frac{\sin \left(\frac{\omega }{\sqrt{2} \sqrt{E}}\right)}{\pi  \hbar \omega }+\mathcal{O}(\hbar)
\end{equation}
We have demonstrated that the Bessel kernel reduces to the sine kernel $K_{\mathrm{sin}}=\frac{\sin \left(\pi \rho_{0}(E)\omega\right)}{\pi  \omega}$ in the $\tau$-scaling limit, where we identify the disk-level Bessel spectral density $\rho_{\mathrm{Bessel}}(E)=\frac{1}{\pi \sqrt{2E}}$ \footnote{One can also obtain the full spectral density $\rho(E)=\frac{J_0\left(\frac{\sqrt{2E}}{\hbar}\right){}^2+J_1\left(\frac{\sqrt{2E} }{\hbar}\right){}^2}{2 \hbar^2}$, computed by plugging \eqref{bess_eigen} into \eqref{CD_kernel}. By performing the limit $\hbar\rightarrow 0$ and using once again the asymptotic expansion for the Bessel functions, we recover the disk-level spectral density. }. We note it exactly matches with the low-energy limit of the SJT disk spectral density \eqref{disk}, once we divide the latter by a factor of 2 to switch from the bulk to the matrix model notation \footnote{In fact each energy eigenvalue for SJT is double-dengerate.}.

Thus, the genus-zero contribution \footnote{With genus-zero contribution we mean that we are moltiplying by a factor of $\hbar=e^{S_{0}}$ in order to isolate the leading term after $\tau$-scaling. }to the spectral form factor for the Bessel model in the $\tau$-scaling limit yields:
\begin{equation}\label{genus-zero}
\mathrm{SFF}_{\mathrm{Bessel}}=\lim_{\hbar \to 0} \hbar \langle Z\left(\beta + i\tau/\hbar\right) Z\left(\beta - i\tau/\hbar\right) \rangle = \int_{0}^{\infty} \mathrm{d} E e^{-2 \beta E} \min\left(\frac{1}{\pi \sqrt{2E}}, \frac{\tau}{2\pi}\right)
\end{equation}
where we performed the integral over $\omega$ in \eqref{two-point2}, which  as $\hbar \rightarrow 0$ is just given by the Fourier transform of $K_{\mathrm{sin}}^2$. 

We can actually go beyond the Bessel model approximation and argue that the structure \eqref{genus-zero} is still valid for the full SJT gravity as well. In fact we saw in \eqref{measure} that, far from the spectral edge, the Altland-Zirnbauer $(1, 2)$ matrix model is governed by the same integral measure characterizing matrix models of unitary type. Interestingly, when $E_1-E_2 \sim \mathcal{O}(\hbar)$, it turns out the two-body eigenvalue correlator is universal for all unitary matrix models \cite{Brezin:1993qg, Dyson:1962es}, the repulsion between close energy levels being always captured by the sine kernel in this limit. As a consequence, we will argue any possibile deviation from the universal sine kernel for the $(1, 2)$ Altland-Zirnbauer might only emerge as one approaches the spectral edge, a regime where the potential is wildly varying. However, we have just proven in \eqref{Bessel_sin} that in the $\tau$-scaling limit the sine kernel behavior keeps holding for the Bessel model, which indeed probes the regime of SJT near the spectral edge.

\section{Exact $\tau$-scaled SFF for the simplified model}\label{exact}

In this Appendix we derive the exact form of the $\tau-$scaled SFF for the toy model of section \ref{simplified} and we compare the result obtained with the trans-series \eqref{bomber}. 

We recall that the spectral density of this model is $\rho_{\mathrm{LE}}(E)=\frac{1+\alpha E}{2 \pi \sqrt{E}}$, therefore according to the discussion of section \ref{Bessel} its $\tau-$scaled SFF admits the following integral representation:
\begin{equation} 
\begin{split}
\mathrm{SFF}_{\mathrm{LE}} & = \int_{0}^{\infty} \mathrm{d} E e^{-2 \beta E} \min\left(\frac{1+\alpha E}{2 \pi \sqrt{E}}, \frac{\tau}{2\pi}\right) = \\
& = \frac{\tau}{4 \pi \beta} - \theta(\tau - 2 \sqrt{\alpha}) \int_{\mathrm{E}_1(\tau)}^{\mathrm{E}_2(\tau)} \mathrm{d} E e^{-2 \beta E} \left( \frac{\tau}{2\pi} - \frac{1+\alpha E}{2 \pi \sqrt{E}}  \right) ,
\end{split}
\end{equation}
where in the second line we used the identity $\min(x,y) = x - (x-y) \theta (x-y)$ and we denoted with $\mathrm{E}_{1,2}^*(\tau)$ the solutions to the equation $\rho_{\mathrm{LE}}(E) = \frac{\tau}{2 \pi}$ when they exist (i.e. when $\tau > 2 \sqrt{\alpha}$).

Unlike the case of SJT gravity, $\mathrm{E}_{1,2}^*(\tau)$ can be readily obtained in this model and they have the following form: 
\begin{equation}
    \mathrm{E}_{1}^*(\tau) = \frac{\tau  \left(\tau -\sqrt{\tau ^2-4 \alpha }\right)-2 \alpha }{2 \alpha ^2} , \quad  \mathrm{E}_{2}^*(\tau) = \frac{\tau  \left(\tau + \sqrt{\tau ^2-4 \alpha } \right)-2 \alpha }{2 \alpha ^2}
\end{equation}
Using these expressions and observing that 
\begin{equation}
    \int \mathrm{d} E e^{-2 \beta E} \left( \frac{\tau}{2\pi} - \frac{1+\alpha E}{2 \pi \sqrt{E}}  \right) = -\frac{4 \sqrt{\beta } e^{-2 \beta  E} \left(\tau -\alpha  \sqrt{E}\right)\!+\!\sqrt{2 \pi } (\alpha +4 \beta ) \text{erf}\left( \sqrt{2\beta E}\right)}{16 \pi  \beta ^{3/2}} +C,
\end{equation}
it is immediate to arrive at the exact result for $\tau-$scaled SFF:
\begin{equation}
    \begin{split}
& \mathrm{SFF}_{\mathrm{LE}}  =   \frac{\tau}{4 \pi \beta} - \frac{1}{16 \pi  \beta ^{3/2}} \theta(\tau - 2 \sqrt{\alpha}) \left[4 \sqrt{\beta} e^{-2 \beta \mathrm{E}_{1}^*(\tau)} \left(\tau - \alpha \sqrt{\mathrm{E}_{1}^*(\tau)} \right) + \right.\\
& -\! \left. 4 \sqrt{\beta} e^{-2 \beta \mathrm{E}_{2}^*(\tau)} \left(\tau - \alpha \sqrt{\mathrm{E}_{2}^*(\tau)} \right)\! +\! \sqrt{2 \pi} (\alpha + 4 \beta) \left( \text{erf}\left(\sqrt{2 \beta \mathrm{E}_{1}^*(\tau) } \right) - \text{erf}\left(\sqrt{2 \beta \mathrm{E}_{2}^*(\tau) } \right)\! \right)\! \right]
\end{split}
\end{equation}
This formula looks quite different from the trans-series \eqref{bomber} of the main text, however it can be recast in a form more similar to \eqref{bomber} by performing an expansion around $\tau = +\infty$.

We observe that when $\tau \to + \infty$, $\mathrm{E}_{1}^*(\tau) \sim \left( \frac{1}{\tau} \right)^2$ and $\mathrm{E}_{2}^*(\tau) \sim \frac{\tau^2}{\alpha^2}$, therefore $\mathrm{SFF}_{\mathrm{LE}}$ consists of two pieces in this limit: one piece coming from $\mathrm{E}_{1}^*(\tau)$ which analytical around $\tau= + \infty$  and the other coming from $\mathrm{E}_{2}^*(\tau)$ which is non-analytical.\footnote{Actually, the constant term of \eqref{expansion} comes from the expansion 
 of $\text{erf}\left(\sqrt{2 \beta \mathrm{E}_{2}^*(\tau)} \right) $ and it is the only analytical term  pertaining to a function of $\mathrm{E}_{2}^*(\tau)$.} Explicitly:
\begin{equation} \label{expansion}
\begin{split}
   \mathrm{SFF}_{\mathrm{LE}} & = \frac{\alpha +4 \beta }{8 \left(\sqrt{2 \pi } \beta ^{3/2}\right)}-\frac{1}{2 \pi  \tau }+\frac{\beta -2 \alpha }{6 \pi  \tau ^3}+O\left(\frac{1}{\tau^5 }\right) +\\
   & - \left(\frac{\alpha ^2 e^{\frac{4 \beta }{\alpha }}}{16 \pi  \beta ^2 \tau }-\frac{\alpha ^2 e^{\frac{4 \beta }{\alpha }} \left(\alpha ^2-8 \beta ^2\right)}{64 \left(\pi  \beta ^3\right) \tau ^3} +O\left(\frac{1}{\tau^5 }\right)\right) e^{ -\frac{2 \beta  \tau ^2}{\alpha ^2}} 
\end{split}
\end{equation}
By comparison we can verify that the first line of \eqref{expansion} perfectly matches with the large $\tau$ expansion of the trans-series in \eqref{bomber} obtained in the main text, and we checked this for the first $15$ terms. Instead, the second line features a non analytical term in $\alpha$ around $\alpha=0$ and can not be captured by the methods of section \ref{simplified}. The reason for this fact is that in \eqref{binomial} we performed a Taylor expansion around $\alpha=0$ which for its nature can not reproduce non-analytical terms such as $e^{ -\frac{2 \beta  \tau ^2}{\alpha ^2}} $.

\section{More detailed calculations for SJT gravity}\label{degregori}
In this Section we explicitly compute the general derivatives that are needed in the main text. The fundamental mathematical tool that we will use is the the Faa di Bruno's formula, a generalization of the chain rule for higher order derivatives: 
\begin{equation}
\frac{\text{d}^n}{\text{d}x^n} f(g(x)) = \sum_{k=0}^{n}f^{(k)}(g(x)) \ B_{n, k}(g'(x), g''(x), ..., g^{(n-k+1)})
\end{equation}
where $B_{n, k}( x_1, x_2, ...,  x_{n-k+1})$ are exponential Bell polynomials. Many properties are known for these polynomials but for our purposes it is sufficient to know the two following relations:
\begin{equation}
\begin{split}
B_{n, k}(ab\, x_1, ab^2\, x_2, ..., ab^{n-k-1}\, x_{n-k+1})&=a^k b^n  B_{n, k}( x_1, x_2, ...,  x_{n-k+1}) \\
  B_{n, k}(1, 1, ..., 1)&=S(n, k)
\end{split}
\end{equation}
The last ingredient we are missing is the following derivative of the incomplete Gamma function and generalized Laguerre polynomial:
\begin{equation}
\frac{\de^n}{\de z^n} \Gamma(a, z)= z^{-n}\sum_{s=0}^{n}(-1)^n \binom{n}{s} \frac{\Gamma(-a+s)}{\Gamma(-a)} \Gamma(a-s+n, z), \quad \frac{\partial ^mL_n^{\lambda }(z)}{\partial z^m}=(-1)^m L_{n-m}^{m+\lambda }(z)
\end{equation}
Exploiting the above relations, one can write down the following master formulae:
\begin{equation}\label{identity1}
\begin{split}
&\lim_{\nu\rightarrow 0}\frac{\text{d}^{2n-k}}{\text{d}\nu ^{2n-k}} \left( \Gamma \left(n+1,\frac{e^{- \nu }2\beta}{\tau^2}\right) \right) =\\
&=\sum _{b=0}^{2 n-k} (-1)^{k+b} S(2n-k, b) \sum _{c=0}^b (-1)^c\binom{b}{c}\frac{  \Gamma (n+2)  }{\Gamma (n+2-c)} \Gamma \left(b-c+n+1,\frac{2 \beta }{\tau ^2}\right)
\end{split}
\end{equation}
\begin{equation}\label{identity2}
\begin{split}
&\lim_{\nu\rightarrow 0}\frac{\text{d}^{2n-k}}{\text{d}\nu ^{2n-k}} \left( e^{-\frac{\nu}{2}}\Gamma \left(n+\frac12,\frac{e^{- \nu }2\beta}{\tau^2}\right) \right) =\\
&\frac{\Gamma \left(n+\frac{3}{2}\right) }{(-2)^{2n-k}}\sum _{s=0}^{2 n-k} 2^{s} \binom{2 n-k}{s} \sum _{b=0}^s  (-1)^b S(s,b) \sum _{c=0}^b (-1)^c \frac{\binom{b}{c}  }{\Gamma \left(n+\frac{3}{2}-c\right)} \ \Gamma \left(b-c+n+\frac{1}{2},\frac{2 \beta }{\tau ^2}\right)
\end{split}
\end{equation}
\begin{equation}\label{identity3}
\begin{split}
& \lim_{\nu \to 0}\frac{\mathrm{d}^{2n-k}}{\mathrm{d}\nu^{2n-k}} \left(e^{-\nu(n+1)} e^{-\frac{2\beta}{\tau^2}e^{-\nu}} \sum _{m=0}^l (-1)^m \binom{l}{m} L_{l-1}^{\left\lfloor \frac{m}{2}\right\rfloor +n-l+1 }\left(\frac{2 \beta  }{\tau ^2}e^{-\nu }\right)  \right)=e^{-\frac{2 \beta  }{\tau ^2}} \times\\
&\sum _{t=0}^{2 n-k}A_{n,k,t} \sum _{b=0}^t  \binom{t}{b}     B_b\left(-\frac{2 \beta }{\tau ^2}\right)\sum _{r=0}^{t-b} \left(-\frac{2 \beta }{\tau ^2}\right)^{r}\! S(t-b, r) \sum _{m=0}^l (-1)^m \binom{l}{m} L_{l-r-1}^{n-l+r+\left\lfloor \frac{m}{2}\right\rfloor +1}\!\!\left(\frac{2 \beta }{\tau ^2}\right) 
\end{split}\end{equation}
with $A_{n,k,t}=\frac{\binom{2 n-k}{t}(-1)^k}{\left(n+1\right)^{k-2n+t}}$ and $B_n\left(x\right)$ the Bell polynomial.
We will also need the following identities:
\begin{equation}\label{identity4}
\begin{split}
\Gamma \left(n+\frac{1}{2},z\right)&=\text{erfc}\left(\sqrt{z}\right) \Gamma \left(n+\frac{1}{2}\right)+(-1)^{n-1} e^{-z}\sqrt{z} \sum _{s=0}^{n-1} (-z)^s \left(\frac{1}{2}-n\right)_{-s+n-1} \\
\Gamma \left( n, z \right) &= (n-1)! e^{-z} \sum_{s=0}^{n-1}\frac{z^{s}}{s!}
\end{split}
\end{equation}
We will now manipulate the expression \eqref{bomber2} of the main text.
First of all, exploiting the properties of the Stirling numbers of the second kind $S(k,0)=\delta_{k, 0}$ and $S(k,1)=1-\delta_{k, 0}$, we reorganize the transeries as:
\begin{equation}\label{bomber3}
\begin{split}
\mathrm{SFF}_{\mathrm{SJT}}(\beta,\tau)&\!=\!\frac{\tau}{4 \pi \beta}\!+\!\lim_{\nu \to 0}\left\{\!\sum_{n=0}^{\infty} \frac{(4\pi)^{2n}}{(2n)!}   \ \frac{\text{d}^{2n} \tilde{\mathcal{I}}_{n,0} (\nu)}{\text{d}\nu ^{2n}}  \! -\frac12\sum_{n=1}^{\infty} \frac{(4\pi)^{2n}}{(2n)!}   \sum_{k=1}^{2n} \binom{2n}{k}    \frac{\text{d}^{2n-k}\tilde{\mathcal{I}}_{n,1} (\nu)}{\text{d}\nu ^{2n-k}}   +\right.\\
&\left. +\sum_{n=1}^{\infty} \frac{(4\pi)^{2n}}{(2n)!} \sum_{k=2}^{2n} \binom{2n}{k}    \frac{\text{d}^{2n-k}}{\text{d}\nu ^{2n-k}} \left(\sum_{l=2}^{k}(-2)^{-l}  S(k, l)  \tilde{\mathcal{I}}_{n,l} (\nu) \right)\right\}
\end{split}
\end{equation}
Thanks to the identities \eqref{identity1}, \eqref{identity2} and \eqref{identity4}, one can write:
\begin{equation}\label{iniesta}
\begin{split}
&\lim_{\nu\rightarrow 0} \ \frac{\text{d}^{2n}}{\text{d}\nu ^{2n}}   \tilde{\mathcal{I}}_{n,0} (\nu)=\\
& =\frac{1}{\pi 2^{2n+1}} \left(\frac{1}{2 \beta }\right)^{n+\frac{1}{2}} \Gamma \left(n+\frac{1}{2}\right) \text{erfc}\left(\frac{\sqrt{2 \beta }}{\tau }\right)+\frac{e^{-\frac{2 \beta }{\tau ^2}}}{\pi \tau}\frac{1}{(2\beta)^n}\left[\mathcal{T}_{n}\left(\frac{2\beta}{\tau^2}\right)+\mathcal{S}_{n}\left(\frac{2\beta}{\tau^2}\right)\right]
\end{split}
\end{equation}
where $\mathcal{T}_{n}\left(\frac{2\beta}{\tau^2}\right)$ and $\mathcal{S}_{n}\left(\frac{2\beta}{\tau^2}\right)$ are polynomials in $\frac{2\beta}{\tau^2}$ of degree $3n-1$, defined by:
\begin{equation}\label{iniesta2}
\begin{split}
\mathcal{S}_{n}\left(\frac{2\beta}{\tau^2}\right) &=\frac{\Gamma (n+2)}{2} \sum _{b=0}^{2 n} (-1)^{b+1} S(2n,b)\sum _{c=0}^b (-1)^c \binom{b}{c}\frac{  \Gamma(b-c+n+1) }{\Gamma (-c+n+2)}\sum _{t=0}^{b-c+n} \frac{1}{t!} \left(\frac{2 \beta }{\tau ^2}\right)^{t-1} \\
\mathcal{T}_{n}\left(\frac{2\beta}{\tau^2}\right) &=\mbox{\small $\displaystyle\frac{(-1)^{n+1}  }{2^{2n+1}} \sum _{s=0}^{2 n} 2^s \binom{2 n}{s} \sum _{b=0}^s S(s,b) \sum _{c=0}^b \frac{\binom{b}{c}\Gamma \left(n+\frac{3}{2}\right) }{\Gamma \left(n-c+\frac{3}{2}\right)} \!\!\sum _{t=0}^{b-c+n-1}\!\! \left(-\frac{2 \beta }{\tau ^2}\right)^{\!t} \!\!\frac{\Gamma \left(-t-\frac{1}{2}\right)}{\Gamma \left(\frac{1}{2}-c-b-n\right)}$}
\end{split}
\end{equation}
Thanks to the identities \eqref{identity2} and \eqref{identity4}, one can then write:
\begin{equation}\label{bobovieri}
-\frac12\lim_{\nu\rightarrow 0}\sum_{k=1}^{2n} \binom{2n}{k}   \ \frac{\text{d}^{2n-k}}{\text{d}\nu ^{2n-k}}   \tilde{\mathcal{I}}_{n,1} (\nu)=\frac{e^{-\frac{2\beta}{\tau^2}}}{\pi \tau} \frac{1}{(2\beta)^n}\mathcal{Q}_{n}\left(\frac{2\beta}{\tau^2}\right)
\end{equation}
where $\mathcal{Q}_{n}\left(\frac{2\beta}{\tau^2}\right)$ is a polynomial in $\frac{2\beta}{\tau^2}$ of degree $3n-2$, whose explicit form is given by:
\begin{align}
\label{bobovieri2}
\mathcal{Q}_{n}\left(\frac{2\beta}{\tau^2}\right)&=\frac{(-1)^{n+1} \Gamma \left(n+\frac{3}{2}\right)}{4} \sum_{k=1}^{2n} \frac{\binom{2n}{k} }{(-2)^{2n-k}}\sum _{s=0}^{2 n-k} 2^{s} \binom{2 n-k}{s}
\times\\ &\times \sum _{b=0}^s  S(s,b) \sum _{c=0}^b \frac{\binom{b}{c}  }{\Gamma \left(n+\frac{3}{2}-c\right)} \sum _{t=0}^{b-c+n-1} \left(-\frac{2\beta}{\tau^2}\right)^t \left(\frac{1}{2}-n-b+c\right)_{b-t-c+n-1}\nonumber
\end{align}
Finally, thanks to the identities \eqref{identity3} and \eqref{identity4}, we arrive at:
\begin{equation}\label{nesta}
\begin{split}
&\lim_{\nu \rightarrow 0} \sum_{k=2}^{2n} \binom{2n}{k}   \ \frac{\text{d}^{2n-k}}{\text{d}\nu ^{2n-k}} \left(\sum_{l=2}^{k}(-2)^{-l}  S(k, l)  \tilde{\mathcal{I}}_{n,l} (\nu) \right)=\frac{e^{-\frac{2 \beta }{\tau ^2}}}{\pi \tau}\frac{1}{(2\beta)^n} \mathcal{P}_{n}\left(\frac{2\beta}{\tau^2}\right)
\end{split}
\end{equation}
where $\mathcal{P}_{n}\left(\frac{2\beta}{\tau^2}\right)$ is a polynomial in $\frac{2\beta}{\tau^2}$ of degree $3n-2$, expressed as:
\begin{equation}
\begin{split}
\mathcal{P}_{n}\left(\frac{2\beta}{\tau^2}\right)&=(-1)^{n+1} \ \sum_{k=2}^{2n} \binom{2n}{k} \sum_{l=2}^{k} \frac{S(k, l) (l-2)!}{2^{l}}\sum _{t=0}^{2 n-k}A_{n,k,t} \sum _{b=0}^t  \binom{t}{b}     B_b\left(-\frac{2 \beta }{\tau ^2}\right) \times  \\
& \times \sum _{r=0}^{t-b} \left(-\frac{2 \beta }{\tau ^2}\right)^{r+n} S(t-b, r)\sum _{m=0}^l (-1)^m \binom{l}{m} L_{l-r-1}^{-l+n+r+\left\lfloor \frac{m}{2}\right\rfloor +1}\left(\frac{2 \beta }{\tau ^2}\right) 
\end{split}
\end{equation}
Because of the common structure of \eqref{iniesta}, \eqref{bobovieri} and \eqref{nesta}, it is useful to introduce the polynomial:
\begin{equation}\label{Z}
\begin{split}
\mathcal{Z}_{n}\left(x\right)=\mathcal{T}_{n}\left(x\right)+\mathcal{S}_{n}\left(x\right) +  \mathcal{Q}_n\left(x\right) + \mathcal{P}_{n}\left(x\right)
\end{split}
\end{equation}
through which one can arrive at the result \eqref{bomber4} of the main text. Below, we list the first few polynomials $\mathcal{Z}_{n}\left(x\right)$:
\begin{align}\label{Z1}
\mathcal{Z}_{1}\left(x\right)&=\frac18 \qquad \mathcal{Z}_{2}\left(x\right)=-\frac{3 x^2}{16}+\frac{x}{32}+\frac{3}{64} \qquad \mathcal{Z}_{3}\left(x\right)=\frac{15 x^4}{32}-\frac{105 x^3}{64}\!+\!\frac{x^2}{128}\!+\!\frac{5 x}{256}+\frac{15}{512} \nonumber\\
\mathcal{Z}_{4}\left(x\right)&=-\frac{105 x^6}{64}+\frac{1995 x^5}{128}-\frac{7623 x^4}{256}+\frac{x^3}{512}+\frac{7 x^2}{1024}+\frac{35 x}{2048}+\frac{105}{4096} \\
\mathcal{Z}_{5}\left(x\right)&\!=\!\frac{945 x^8}{128}-\frac{34335 x^7}{256}\!+\!\frac{351855 x^6}{512}-\frac{986205 x^5}{1024}\!+\!\frac{x^4}{2048}\!+\!\frac{9 x^3}{4096}\!+\!\frac{63 x^2}{8192}\!+\!\frac{315 x}{16384}\!+\!\frac{945}{32768}\nonumber
\end{align}
We notice that, although $\mathcal{T}_{n}\left(x\right)$, $\mathcal{S}_{n}\left(x\right)$, $\mathcal{Q}_n\left(x\right)$ and  $\mathcal{P}_{n}\left(x\right)$ have degree at least $3n-2$, the resulting polynomial $\mathcal{Z}_{n}\left(x\right)$ has a degree $2n-2$, because of highly non trivial cancellations occurring among the constituent polynomials. Although we have determined in principle a close analytical form for $\mathcal{Z}_{n}$, it is quite intricated. Because of that,  we have empirically analyzed  the structure of the first $20$ polynomials and we have been able to partially determine a more manageable expression for $\mathcal{Z}_{n}$, i.e.
\begin{equation}\label{Z2}
\begin{split}
\mathcal{Z}_{n}\left(x\right)&=\sum _{k=0}^{n-1} \frac{(2 n-1)\text{!!} x^k}{(2 k+1)\text{!!} 2^{3 n-k}}+\sum_{k=n}^{2n-3} \frac{(-1)^{k+n+1} \ a_{k}}{2^{3n-k}}x^{k}+\left(1-\delta_{n,1}\right)\frac{(-1)^{n+1}(2n-1)!!}{2^{n+2}} x^{2n-2},
\end{split}
\end{equation}
where $a_{k}$ are coefficients that we have not been able to fix in a direct way but can be deduced from \eqref{iniesta}, \eqref{bobovieri} and \eqref{nesta}.

We finally point out that, by exploiting the form \eqref{Z2} for the polynomials $Z_{n}$, it is also possible to rewrite the transeries expansion \eqref{bomber33} of the main text in another useful form:
\begin{equation}
\begin{split}
\mathrm{SFF}_{\mathrm{SJT}}(\beta,\tau)\!=\!&\frac{\tau}{4 \pi \beta}-\frac{\tau}{4 \pi \beta} e^{-\frac{2\beta}{\tau^2}}+\frac{1}{2 \sqrt{2 \pi \beta}}\!\left[\mathrm{erfc} \left(\frac{\sqrt{2\beta}}{\tau}\right)+\sum_{n=1}^{+\infty} \frac{\left(\frac{\pi^2}{2 \beta}\right)^n}{n!} \frac{\Gamma \left(\frac12 +n, \frac{2\beta}{\tau^2}\right)}{\Gamma \left(\frac12 +n\right)}\right]\!+\!\mathcal{R},
\end{split}
\end{equation}
where we define $\mathcal{R}$ as: 
\begin{equation}
\begin{split}
\mathcal{R}=\frac{e^{-\frac{2\beta}{\tau^2}}}{\pi \tau} \sum_{n=1}^{+\infty} \frac{\left(\frac{4 \pi}{\sqrt{2\beta}}\right)^{2n}}{(2n)!}\sum_{k=n}^{2n-3} \frac{(-1)^{k+n+1}}{2^{3n-k}} a_{k} \left(\frac{2\beta}{\tau^2}\right)^{k}+\frac{\tau^3}{16 \pi \beta^2} e^{-\frac{2\beta}{\tau^2}} \left(1-e^{-\frac{8 \pi^2 \beta}{\tau^4}}\right)-\frac{\pi  e^{-\frac{2 \beta }{\tau ^2}}}{2 \beta  \tau }.
\end{split}
\end{equation}

\newpage

\bibliography{biblio}
\end{document}